\documentclass[fleqn,usenatbib]{mnras}

\usepackage{newtxtext,newtxmath}
\usepackage{natbib}
\usepackage{multicol}
\usepackage{xcolor}

\usepackage[T1]{fontenc}
\usepackage{ae,aecompl}

\usepackage{graphicx}	
\usepackage{amsmath}	

\usepackage{amssymb}	
\usepackage{bm}

\usepackage{comment}

\newcommand{\maxPram}{P_\mathrm{ram,\,max}}
\newcommand{\intPram}{P_\mathrm{ram,\,int}}
\newcommand{\quenchPram}{P_\mathrm{ram,\,strip}}
\newcommand{\impulsePram}{I_\mathrm{ram}}
\newcommand{\periPram}{P_\mathrm{ram,\,peri}}
\newcommand{\MHI}{M_\mathrm{HI}}
\newcommand{\rstarhalf}{R_{*,1/2}}
\newcommand{\mystar}{*}
\newcommand{\Msun}{\mathrm{M}_{\odot}}
\newcommand{\Mstar}{M_{\mystar}}
\newcommand{\Mhalo}{M_\mathrm{halo}}
\newcommand{\Mtwohm}{M_\mathrm{200m}}
\newcommand{\lcdm}{\Lambda\mathrm{CDM}}
\newcommand{\rtwoh}{R_{\rm 200m}}
\newcommand{\vmax}{V_\mathrm{max}}
\newcommand{\mpc}{\mathrm{Mpc}}
\newcommand{\Mpc}{\textnormal{Mpc}}
\newcommand{\kpc}{\mathrm{kpc}}

\newcommand{\degree}{^\circ}
\newcommand{\gyr}{\mathrm{Gyr}}
\newcommand{\myr}{\mathrm{Myr}}
\newcommand{\dhost}{d_{\rm host}}

\definecolor{jscolor}{HTML}{067C09}

\title[A jolt to the system]{A jolt to the system: ram pressure on low-mass galaxies in simulations of the Local Group}

\author[J. Samuel et al.]
{Jenna Samuel$^{1}$\thanks{E-mail: jenna.samuel@austin.utexas.edu}\thanks{NSF Astronomy and Astrophysics Postdoctoral Fellow},
Bhavya Pardasani$^{2}$,
Andrew Wetzel$^{3}$,
Isaiah Santistevan$^{2}$,
Michael Boylan-Kolchin$^{1}$,
\newauthor
Jorge Moreno$^{4,5}$,
Claude-Andr{\'e} Faucher-Gigu{\`e}re$^{6}$
\\
$^{1}${Department of Astronomy, The University of Texas at Austin, 2515 Speedway, Stop C1400, Austin, TX 78712, USA}\\
$^{2}$Department of Astronomy, University of Illinois at Urbana-Champaign, 1002 W Green St, Urbana, IL 61801, USA\\
$^{3}$Department of Physics and Astronomy, University of California, Davis, CA 95616, USA\\
$^{4}$Department of Physics and Astronomy, Pomona College, Claremont, CA 91711, USA\\
$^{5}$Downing College, University of Cambridge, Cambridge CB3 OHA, UK\\
$^{6}${Department of Physics and Astronomy and CIERA, Northwestern University, 2145 Sheridan Road, Evanston, IL 60208, USA}\\
}

\date{Accepted XXX. Received YYY; in original form ZZZ}

\pubyear{2023}

\begin{document}
\label{firstpage}
\pagerange{\pageref{firstpage}--\pageref{lastpage}}
\maketitle

\begin{abstract}
Low-mass galaxies are highly susceptible to environmental effects that can efficiently quench star formation.
We explore the role of ram pressure in quenching low-mass galaxies ($M_{*}\sim10^{5-9}\,\rm{M}_{\odot}$) within 2 Mpc of Milky Way (MW) hosts using the FIRE-2 simulations. 
Ram pressure is highly variable across different environments, within individual MW haloes, and for individual low-mass galaxies over time.
The impulsiveness of ram pressure --- the maximum ram pressure scaled to the integrated ram pressure prior to quenching --- correlates with whether a galaxy is quiescent or star-forming.
The time-scale between maximum ram pressure and quenching is anticorrelated with impulsiveness, such that high impulsiveness corresponds to quenching time-scales $<1$ Gyr.
Galaxies in low-mass groups ($M_\mathrm{*,host}\sim10^{7-9}\,\rm{M}_{\odot}$) outside of MW haloes experience typical ram pressure only slightly lower than ram pressure on MW satellites, helping to explain effective quenching via group pre-processing.
Ram pressure on MW satellites rises sharply with decreasing distance to the host, and, at a fixed physical distance, more recent pericentre passages are typically associated with higher ram pressure because of greater gas density in the inner host halo at late times.
Furthermore, the ram pressure and gas density in the inner regions of Local Group-like paired host haloes is higher at small angles off the host galaxy disc compared to isolated hosts.
The quiescent fraction of satellites within these low-latitude regions is also elevated in the simulations and observations, signaling possible anisotropic quenching via ram pressure around MW-mass hosts.
\end{abstract}





\section{Introduction}

The shallow gravitational potentials of low-mass galaxies ($\Mstar\lesssim10^9\,\Msun$) typically cannot provide sufficient restoring force to retain gas against disruptive interactions with the environment.
This renders low-mass galaxies highly susceptible to rapid environmental quenching, and their star formation can be quenched within $\lesssim2$ Gyr of an interaction like infall into a host halo \citep[e.g.,][]{Akins2021,Jahn2022,Samuel2022,Pan2023}.
In particular, low-mass galaxies may efficiently lose gas and quench through ram pressure stripping due to their motion through an ambient gas medium \citep[e.g.,][]{Gunn1972,Abadi1999,Mayer2006,McCarthy2008,Grcevich2009,Cortese2021,Boselli2022}.

Ram pressure stripping is typically thought to proceed outside-in, whereby gas at the outskirts of a halo is easier to remove because of lower gravitational restoring forces \citep[e.g.,][]{McCarthy2008}.
Recent work has also shown that ram pressure can remove gas from the outskirts of a galaxy while compressing the inner dense gas and causing further star formation \citep[e.g.,][]{Tonnesen2009,Genina2019,Wright2019,Hausammann2019,DiCintio2021}.
However, \citet{Fillingham2015} showed that in order to match the observed quiescent fraction of satellite galaxies in the Local Group (LG), the satellites must have encountered dense, clumpy gas causing high ram pressure and rapid quenching.
This paints a picture of two modes of ram pressure that affect a galaxy's star formation: smooth ram pressure that slowly removes circumgalactic medium (CGM) gas and cuts off fresh gas accretion, and impulsive ram pressure that completely strips a galaxy of its interstellar medium (ISM) and rapidly quenches its star formation.

Studies of ram pressure often focus on the densest environments like massive galaxy clusters and groups, where ram pressure stripping is often identified by the presence of so-called jellyfish galaxies with neutral hydrogen (HI) tails extending opposite their direction of motion \citep{Ebeling2014,Poggianti2017,Yun2019}.
In the LG, the gas-poor nature of most low-mass galaxies close to the MW and M31, compared to more isolated galaxies at similar mass, implies that their gas is efficiently removed by ram pressure \citep[e.g.,][]{Putman2021}.
The short time-scales between infall and quenching for observed and simulated LG galaxies also suggest that a rapid process like ram pressure stripping may be responsible for removing their gas \citep[e.g.,][]{Wheeler2014,Fillingham2015,Wetzel2015a,Simpson2018,Samuel2022}.
\citet{Simons2020} recently confirmed that ram pressure in MW haloes is indeed highly stochastic, i.e., it varies significantly on time-scales as short as $\lesssim$50 Myr using the Figuring Out Gas in GalaxIEs (FOGGIE) simulations.
However, it remains unclear what quantitatively defines the distinct modes of smooth and impulsive ram pressure on low-mass galaxies in the LG, and therefore whether or not ram pressure is their decisive quenching mechanism.

Furthermore, even LG galaxies that are far away from the Milky Way (MW) and Andromeda (M31) may be experiencing environmental quenching via ram pressure.
For example, the Wolf-Lundmark-Melotte (WLM) galaxy (930 and 830 kpc from the MW and M31, respectively) has trailing clouds of HI gas indicative of ram pressure stripping \citep{Yang2022}.
Moreover, the collapse of large scale structures like sheets and filaments can shock heat the intergalactic medium (IGM) at early times ($z=2-5$) and quench otherwise isolated galaxies that interact with the shock front \citep{Benitez2013,Pasha2023}.
Low-mass galaxies from the FIRE simulations also show evidence for environmental quenching out to $\approx1\,\mpc$ from MW-mass hosts \citep{Samuel2022}.
Though some of these isolated quiescent galaxies that are backsplash galaxies that have previously interacted with a massive host \citep{Simpson2018,Benavides2021}, many have rapidly (within $1-2$ Gyr) quenched after interactions with hosts ($M_\mathrm{*,host}\sim10^{7-9}\,\rm{M}_{\odot}$) of low-mass groups \citep{Samuel2022}.
Ram pressure in low-mass groups is a likely culprit for such quenching, but this has yet to be fully quantified with simulations.

In this work, we quantify the conditions necessary for ram pressure to rapidly quench low-mass galaxies using cosmological hydrodynamic simulations of the LG.
In Section~\ref{sec:simulations}, we describe the simulations and how we measure ram pressure histories for each galaxy.
In Section~\ref{sec:results}, we: present ram pressure histories for galaxies in different environments (\ref{subsec:rp_histories}), describe the ram pressure conditions necessary to quench galaxies (\ref{subsec:rp_to_quench}), characterize the strength of ram pressure on MW satellites at pericentre passage (\ref{subsec:rp_at_peri}), and examine effects of angular anisotropy in the host CGM on ram pressure in the inner halo (\ref{subsec:cgm_anisotropy}).
In Section~\ref{sec:discussion}, we discuss the implications of our findings for quenching via ram pressure in the Local Group and beyond.
We summarize our main conclusions in Section~\ref{sec:conclusions}.


\section{Simulations}\label{sec:simulations}

We analyze 536 low-mass galaxies identified at $z=0$ around 14 MW/M31-mass host galaxies from the FIRE simulation project\footnote{\url{https://fire.northwestern.edu/}}.
We chose our sample of galaxies using a stellar mass selection of $\Mstar=10^{5-10}\,\Msun$ and by requiring them to reside within $\leq2\,\mpc$ of a MW/M31-mass host.
These galaxies lie within the high resolution regions of the simulations ($\dhost\lesssim2\,\Mpc$), and inhabit well-resolved dark matter haloes \citep{Samuel2020}.
We require that the halos hosting low-mass galaxies in our sample are not significantly contaminated by low-resolution dark matter particles ($M_\mathrm{low-res}/\Mhalo\leq0.02$) and are not actively disrupting ($M_\mathrm{bound}/\Mhalo > 0.4$).
We further require the average stellar density within the stellar half-mass radius of a low-mass galaxy in our sample to be $\geq10^3\,\Msun/\kpc^3$ to avoid spurious galaxy identification, though this could possibly exclude some ultra-diffuse galaxies.
Our sample contains 7 additional galaxies at $\dhost=1-2\,\Mpc$ (in m12i, m12b, and m12w) that were not previously identified by the earlier analysis pipeline used in \citet{Samuel2022}.

Eight of the MW/M31-mass hosts are isolated from other massive hosts, and the other six are in LG-like pairs \citep{Wetzel2016,GK2019a,Samuel2020}.
We simulated the six paired hosts and one isolated host with baryonic mass resolutions of m$_{\rm baryon,ini}=3500-4200\,\Msun$ (m$_{\rm dm}\approx2\times10^4\,\Msun$), and the other seven isolated hosts have m$_{\rm baryon,ini}=7100\,\Msun$ (m$_{\rm dm}=3.5\times10^4\,\Msun$).
We generated cosmological zoom-in initial conditions using \textsc{MUSIC} \citep{Hahn2011} with flat $\lcdm$ cosmologies that are broadly consistent with cosmological parameters inferred by \citet{PlanckCollaboration2018}.

We ran the simulations with the FIRE-2 implementations of fluid dynamics, star formation, and stellar feedback \citep{Hopkins2018}.
FIRE uses the \textsc{GIZMO} Lagrangian meshless finite-mass (MFM) hydrodynamics solver \citep{Hopkins2015} and gravitational forces are solved using an upgraded version of the $N$-body \textsc{GADGET-3} Tree-PM solver \citep{Springel2005}.
\textsc{GIZMO} enables adaptive hydrodynamic gas smoothing (spatial resolution) based on the local density of gas, while simultaneously conserving mass, energy, and momentum to machine accuracy.

FIRE-2's subgrid model for gas implements a metallicity-dependent treatment of radiative heating and cooling over $10-10^{10}$ K \citep{Hopkins2018}, a cosmic ultraviolet background ($z_{\rm reion}\sim10$) \citep{FaucherGiguere2009}, and turbulent diffusion of metals \citep{Hopkins2016,Su2017,Escala2018}.
Star formation in FIRE-2 occurs in gas that is self-gravitating, Jeans-unstable, cold (T $<10^4$ K), dense ($n>1000$ cm$^{-3}$), and molecular \citep{Krumholz2011}.
Several stellar-feedback processes are also included in FIRE-2 via subgrid models, including Type Ia supernovae, mass loss, photoionization, photoelectric heating, and radiation pressure.
Supernovae are individually time-resolved and the FIRE-2 algorithm for coupling their mechanical feedback to the surrounding gas manifestly conserves mass, energy, and momentum \citep{Hopkins2018a}.

We assigned gas to low-mass galaxies/haloes following \citet{Samuel2022}.
Briefly, gas cells assigned to a galaxy/halo must (1) lie within $2\rstarhalf$ (the stellar half-mass radius) of a halo centre and (2) have a total velocity less than $2\,MAX(\vmax,\sigma)$, where $\vmax$ is the maximum circular velocity of the dark matter halo and $\sigma$ is the velocity dispersion of the dark matter halo.
We calculate the approximate gas surface density of a galaxy by summing the total mass of gas cells whose centers fall within a spherical aperture of radius $2\rstarhalf$ and diving by $\pi \rstarhalf^2$, neglecting its full 3D structure.
Note that our gas assignment roughly corresponds to the ISM of a galaxy rather than the CGM, and there are often brief episodes where a galaxy does not have any assigned gas because stellar feedback both accelerates gas beyond our velocity limit and pushes gas outside of our physical aperture.

These simulations reproduce key elements of the LG satellite population such as the stellar mass function, stellar mass-halo mass relation, radial distance distribution, star formation histories, and aspects of satellite planes \citep{Wetzel2016,GK2019a,GK2019b,Samuel2020,Samuel2021,Santistevan2023}.
Importantly, the low-mass galaxies in our simulations have realistic star formation histories (SFH) and quiescent fractions.
Following \citet{Samuel2022}, we define a low-mass galaxy in the simulations as quiescent if no stars have formed within it for the last $200\,\myr$ and it is assigned $\MHI<10^6\,\Msun$ ($\lesssim140$ gas cells) at $z=0$.
Using this definition, the quiescent fraction of satellite galaxies in the simulations rises sharply at $M_{*}\lesssim10^7\,\rm{M}_{\odot}$, similar to the LG \citep{Samuel2022}.
\citet{GK2019b} also demonstrated that the low-mass galaxies far from MW-mass hosts in the simulations (non-satellites) have extended SFH, as expected for galaxies that experience weak to no environmental influence from the host.
However, we note that there is likely some numerical over-quenching at $\Mstar\lesssim10^6\,\Msun$ \citep{Samuel2022}.
More detail on the gas content, quiescent fraction, and implications of resolution for these galaxies can be found in \citep{Samuel2022}.

\begin{figure}
	\includegraphics[width=\columnwidth]{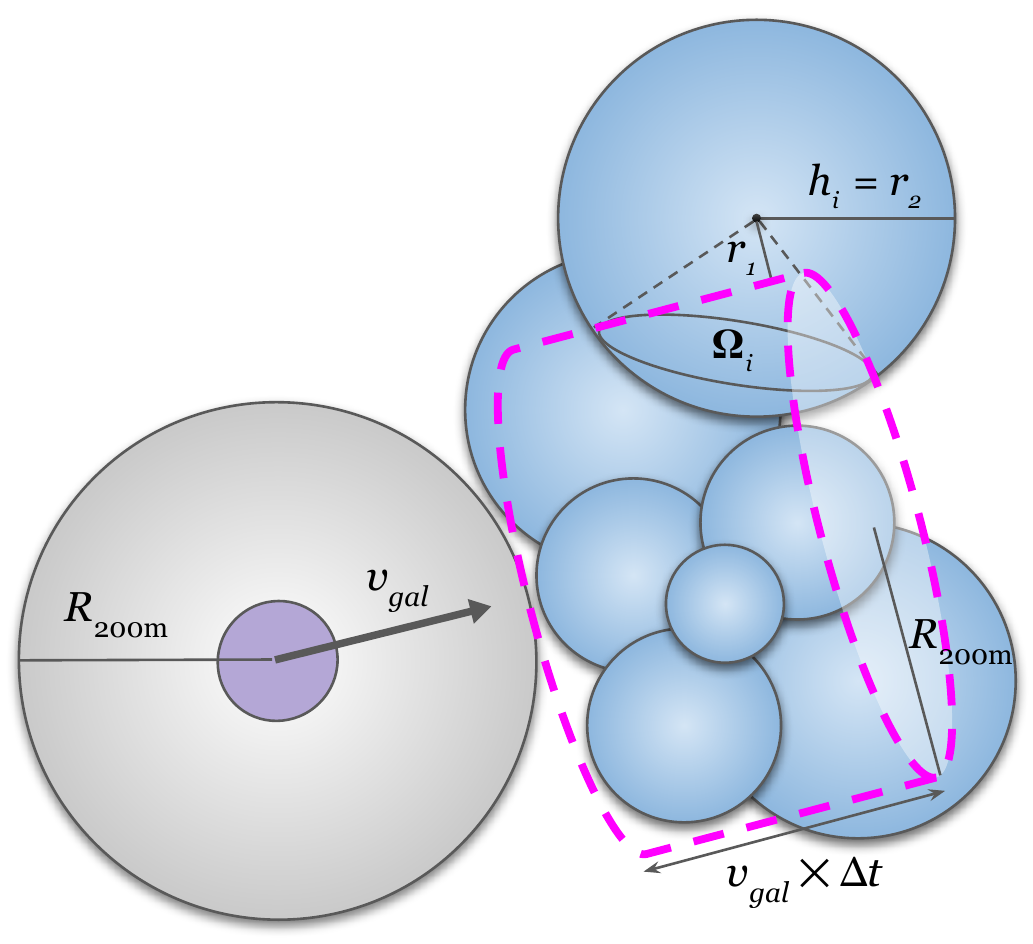}\par
	\vspace{-2 mm}
    \caption{A diagram of the cylindrical region we use to measure the ambient gas density and velocity in our calculation of ram pressure on a low-mass galaxy (at left).
    The cylinder's radius is equal to the incident galaxy's halo radius, and the cylinder's height or length is the distance the galaxy will travel between the current snapshot and the next.
    The purple region within the galaxy's halo represents roughly its stellar extent, $2\rstarhalf$, the distance within which we assign gas to the galaxy.
    The quantities displayed in the upper right illustrate the integration limits for the case where the centre of a gas cell lies outside the cylinder, but its kernel radius overlaps with the cylinder.
    Note that objects are not drawn to scale.
    }
    \label{fig:cyl_diagram}
\end{figure}

\subsection{Measuring ram pressure}

We calculate the localized ram pressure that a low-mass galaxy experiences following \citet{Simons2020}.
We adapt their localized approach to calculating density for our simulations' Lagrangian hydrodynamic scheme, as opposed to spherically-averaging the density profile of the host halo.
This allows us to measure ram pressure at any point within the simulation volume.

We illustrate the set up of our localized measurement in Figure~\ref{fig:cyl_diagram}.
We measure the ambient gas density and velocity relative to a low-mass galaxy within a cylinder in front of the galaxy at each snapshot.
The cylinder's axis points along the direction of the galaxy's velocity, and the cylinder's closest face sits at the galaxy's dark matter halo radius.
The radius of the cylinder is equal to the galaxy's halo radius ($r_\mathrm{cyl} = \rtwoh$).
The height or length of the cylinder is approximately the distance the galaxy will travel between snapshots ($l_\mathrm{cyl} = v_\mathrm{gal} \times \Delta t \approx v_\mathrm{gal} \times 25\,\myr$).
For a galaxy moving at 100 km/s, the length of the cylinder is about 2.5 kpc, which is much smaller than a typical low-mass halo radius ($\sim40$ kpc).

We measure the gas mass within the cylinder by first summing the mass of all gas cells with centre positions that fall inside the cylinder volume:
\begin{equation}\label{eq:gas_inside}
    M_\mathrm{gas,\,inside} \approx \sum_{i}^{} m_{i,\,\mathrm{inside}}.
\end{equation}

In some cases, especially in low density environments, the centre of a gas cell ($\textbf{\textit{x}}_i$) may not lie inside the cylinder, but there is still significant overlap between the gas cell kernel radius ($h_i$) and the cylinder (see, for example, the top right gas cell in Figure~\ref{fig:cyl_diagram}).
We compute this overlapping mass contribution as
\begin{equation}\label{eq:gas_overlap}
    M_\mathrm{gas,\,overlap} \approx \sum_{i}^{} m_{i,\,\mathrm{overlap}} \cdot \mathscr{w}_i,
\end{equation}
by weighting a gas cell's mass by the normalized integral of its kernel density function within the overlapping region,
\begin{equation}\label{eq:integral_weight}
    \mathscr{w}_i =  \frac{\Omega_i\int_{r_1}^{r_2} W(q,h_i) q^2 dq}{4\pi \int_{0}^{h_i} W(q,h_i) q^2 dq}.
\end{equation}

The kernel density function, $W(q,h_i)$, is a cubic spline (see Equation H4 in \citealt{Hopkins2015}) where $q=r/h_i$ is the distance from the gas cell centre normalized by the kernel radius.
We approximate the integration limits as $r_1 = MAX(s_i,z_i)$ and $r_2 = h_i$, where $s_i$ and $z_i$ are the cylindrical distances of the centre of the overlapping gas cell relative to the surface of the cylinder.
We approximate the solid angle of the integral using Equation 2 from \citet{Hopkins2018a},
\begin{equation}\label{eq:solid_angle}
    \Omega_i \approx 2\pi\left(1-\frac{d_i}{\sqrt{d_i^2 + D_\mathrm{cyl,max}^2}}\right).
\end{equation}

This approximates the solid angle subtended by the area of a circle of radius $D_\mathrm{cyl,\,max} = MAX(r_\mathrm{cyl}, l_\mathrm{cyl}/2)$ at a distance $d_i = |\textbf{\textit{x}}_i - \textbf{\textit{x}}_\mathrm{cyl}|$, in other words, the approximate solid angle taken up by the cylinder from the point of view of the overlapping gas cell.
$D_\mathrm{cyl,\,max}$ is typically equal to the cylinder radius, owing to the halo radii of galaxies usually being much larger than the distances the galaxies travel between snapshots.
We divide the total gas mass within the cylinder by the cylinder volume to obtain the average density of ambient gas contributing to ram pressure on a galaxy,
\begin{equation}\label{eq:amb_density}
    \rho_\mathrm{gas,\,ambient} = \frac{1}{V_\mathrm{cyl}} \left( M_\mathrm{gas,\,inside} + M_\mathrm{gas,\,overlap} \right).
\end{equation}

Our integration scheme creates a somewhat deformed cylinder boundary, but it is computationally efficient. 
Our exact choices for cylinder dimensions and integral approximations do not significantly affect our main conclusions, provided that the cylinder is defined outside of the galaxy's halo radius so as to avoid the galaxy's own ISM, which can extend well past $\rstarhalf$ at different times owing to stellar feedback \citep{AnglesAlcazar2017,Hafen2019,Rey2022}.

We calculate ram pressure as
\begin{equation}\label{eq:local_rp}
    P_\mathrm{ram} = \rho_\mathrm{gas,\,ambient} \cdot {v_\mathrm{gal}}^2,
\end{equation}
following \citet[e.g.,][]{Gunn1972}, where $v_\mathrm{gal}$ is the average velocity of gas cells contributing to $\rho_\mathrm{gas,\,ambient}$, relative to the galaxy in question.
Throughout this paper, we present ram pressure in cgs units as ${\rm g \cdot cm^{-1} \cdot s^{-2}}$.

\begin{figure*}
	\includegraphics[width=\textwidth]{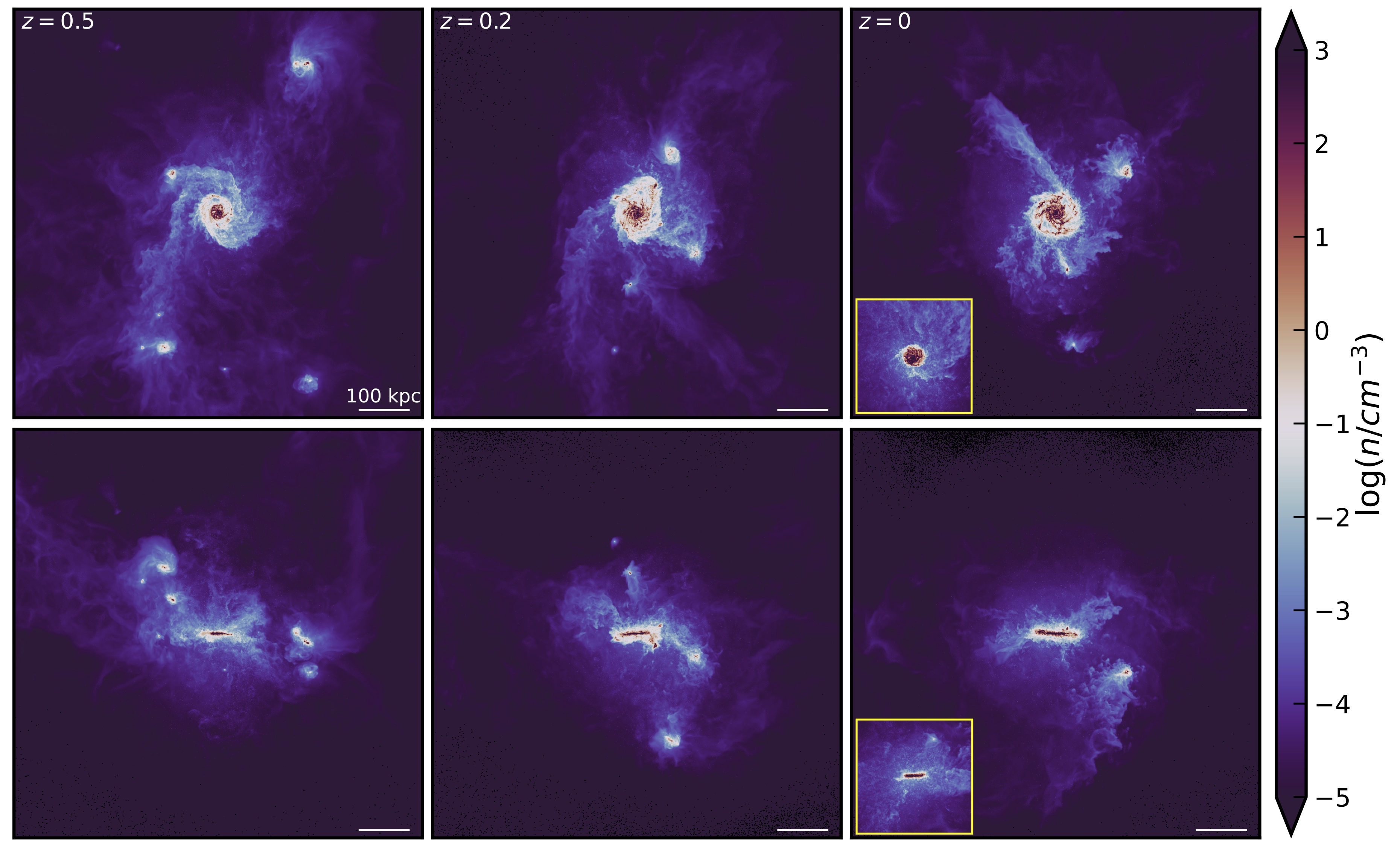}\par
	\vspace{-2 mm}
    \caption{Gas density around a LG-like paired host (Louise) at three different snapshots separated by $\approx2.5$ Gyr over the last $\approx5$ Gyr. 
    The top row is a face-on view of the host and the bottom row is an edge-on view of the host at each snapshot. 
    Gas stripping from satellites is apparent in every panel, and the satellites leave dense gas behind in the host CGM.
    Insets in the right two panels show the gas disc of an isolated host (m12i) at the same scale, demonstrating that the gas discs of the paired hosts extend farther out into the inner host halo and illustrating the anisotropic inner CGM density of paired hosts.
    }
    \label{fig:host_vis}
\end{figure*}

We focus on ram pressure over the last 10 Gyr ($z\lesssim1.65$, 400 snapshots) of the simulations to focus on the times when there is a prominent MW host/progenitor \citep{Santistevan2020} and to exclude reionization and high stochasticity in gas density in the early Universe.
Our time baseline captures important evolutionary moments for MW satellites: median lookback time to infall into the MW halo is about 7 Gyr and median lookback time to quenching is 9.6 Gyr \citep{Santistevan2023}.
We note that only 219 low-mass galaxies out of the 407 total quiescent galaxies actually quench during this time period.
The remaining 188 galaxies that quench at earlier times have a median stellar mass of $\Mstar=2.2\times10^5\,\Msun$ and may quench as a result of reionization or numerical over-quenching due to their low masses.


\section{Results}\label{sec:results}

\subsection{The host CGM}\label{sec:host_cgm}

To understand the general relationship between the gaseous MW host halo and satellite galaxy quenching, we characterize the density of the MW host CGM and its effects on satellites.
We first quantify CGM density as a spherically symmetric function of distance from the host.
We note that this model cannot account for local variations or clumpiness in density, but it is useful to estimate the ram pressure typically felt by satellites.
In particular, we expect the inner host halo to be more dense and hence to more efficiently ram pressure strip and quench satellites with close pericentre passages (see Section~\ref{subsec:rp_at_peri}).
In Section~\ref{subsec:cgm_anisotropy}, we further explore the angular anisotropy of the CGM around LG-like paired hosts.

Figure~\ref{fig:host_vis} shows face-on and edge-on visualizations of gas density around a paired MW-mass host over the last $\approx5$ Gyr.
We created these images by selecting all gas within a cubical region of side length $800\,\kpc$ around the host and colouring each pixel by the maximum density along the line of sight.
The host halo gas density is visually higher in the inner regions ($R\lesssim150\,\kpc$), and it decreases by 1-2 orders of magnitude (dex) at $R\gtrsim200\,\kpc$. 
However, the CGM is also highly structured, with evidence of clumps and filaments throughout, but particularly near the host disc plane.
The several satellite galaxies, and the trails of gas left behind them, temporarily enhance the density of the host CGM in localized areas by up to three dex.
\citet{AnglesAlcazar2017} and \citet{Hafen2019} have previously noted that stripped satellite gas enriches the host CGM, but here we point out that it likely significantly contributes to the ram pressure experienced by nearby satellite galaxies and hence contributes to quenching their star formation.

We visualize this particular host (Louise) in Figure~\ref{fig:host_vis} because it is one of the LG-like paired hosts, which have larger gas discs and exhibit enhanced gas density near the plane of the disc past the stellar disc ($R\approx15$ kpc) and into the halo ($R\lesssim100$ kpc).
The stellar discs of the paired hosts are also on average $15-25$ per cent larger in radius than the isolated hosts, depending on the ages of the stellar populations \citep{GK2018,Bellardini2022}. 
For comparison, we show the gas disc of an isolated host (m12i) at the same scale in the insets in the right panels of Figure~\ref{fig:host_vis}.
The gas disc of the isolated is clearly smaller than that of the paired host in both radius and height.
The size difference between isolated and paired hosts has the important consequence that satellites orbiting close to a paired host feel higher ram pressure than satellites orbiting close to an isolated host, especially near the host disc.
We explore this anisotropic ram pressure further in Section~\ref{subsec:cgm_anisotropy}.

\begin{figure}
    \includegraphics[width=\columnwidth]{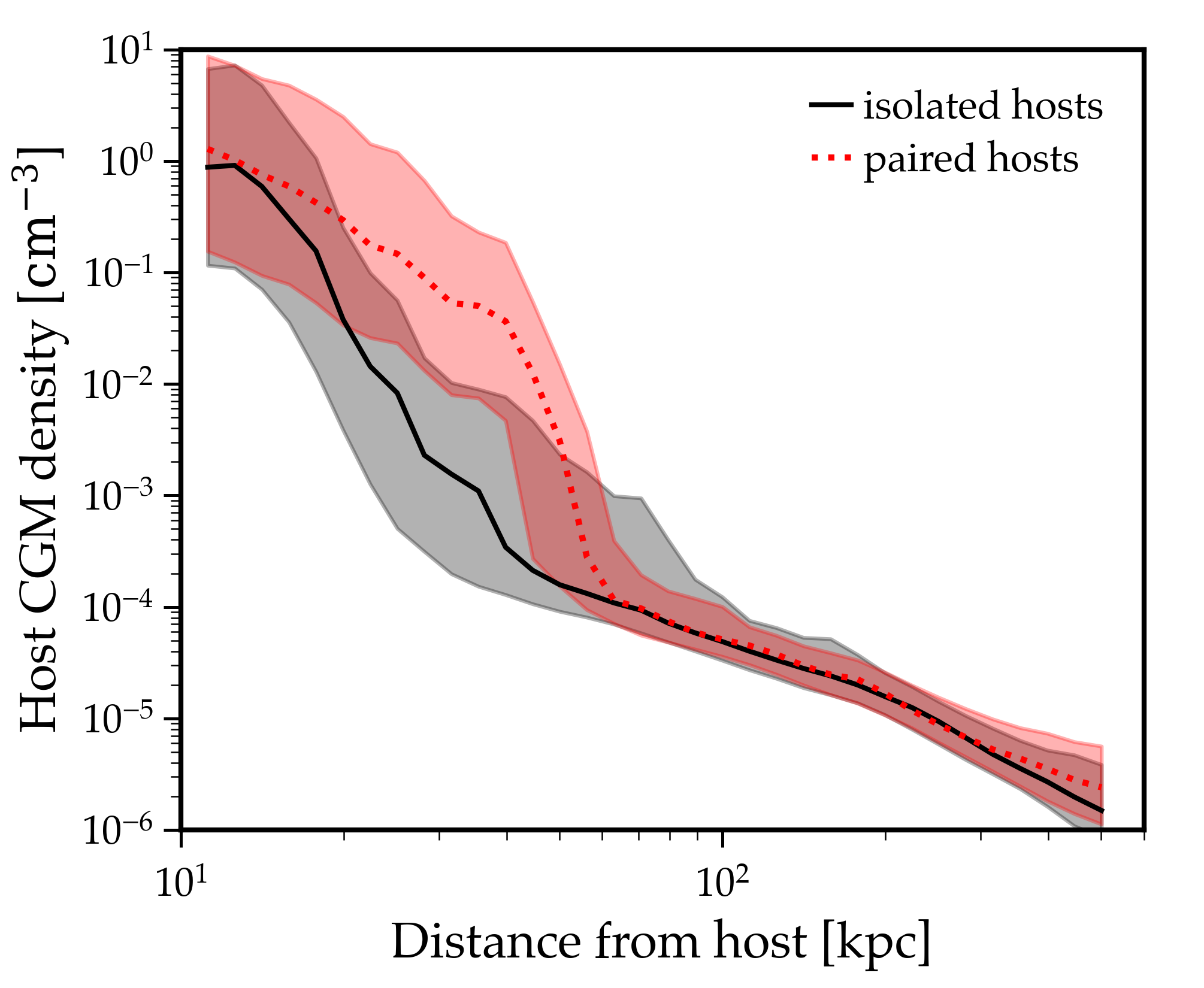}\par
    \vspace{-2 mm}
    \caption{
    The radial profile of CGM gas density around isolated and paired MW-mass hosts.
    We show the host-to-host median and 68 per cent variation in density at $z=0$.
    The median radial profiles of the isolated and paired hosts are nearly identical close to the disc ($\lesssim10$ kpc) and in the outer halo ($\gtrsim100$ kpc), but at $20-50\,\kpc$ the gas around paired hosts is up to two dex more dense than the gas around isolated hosts.
    }
    \label{fig:host_cgm}
\end{figure}

We quantify the differences in the spherically-symmetric CGM profiles of paired and isolated hosts in Figure~\ref{fig:host_cgm} (left).
We computed the median and 68 per cent variation in density within radial shells (35 log-spaced radial bins over $10-500\,\kpc$) around each host.
We show the host-to-host median amongst their median profiles, and the host-to-host median of upper and lower 68 per cent limits on variation in density within individual host halos.
We remove the satellite gas before calculating the CGM density to remove contamination from satellite ISM or recently stripped satellite gas. 
To exclude satellite gas from our measurements of the host halo, we identify the satellites present within the host halo using the halo catalogs and ignore any gas within $10\,R_{*,1/2}$ and $10\,\sigma$ of a satellite 
This is similar to our method for gas assignment described in Section~\ref{sec:simulations}, but we use a more liberal selection to omit recently stripped gas.

Figure~\ref{fig:host_cgm} (left) shows that the host halo gas density at $z=0$ varies over about seven dex and rises significantly at $\lesssim50\,\kpc$.
Notably, between about 20 and 50 kpc, the median density around paired hosts is up to two dex higher than density around isolated hosts.
We explore this extended region of high density around the paired hosts further in Section~\ref{subsec:cgm_anisotropy}.
Beyond $\sim100\,\kpc$, the paired and isolated profiles are quite similar and have much smaller scatter.
Though not shown here, we also note that for half of the hosts the median density in the inner CGM ($\lesssim100\,\kpc$,) has increased by about 1 dex over the last 10 Gyr.

Hydrodynamic interactions like ram pressure are subject to resolution effects in simulations.
Due to the Lagrangian nature of the hydrodynamic solver that we use, the gas cell size (hydrodynamic smoothing kernel radius) scales with gas density, such that regions of higher density are automatically better resolved.
In Appendix~\ref{sec:resolution}, we show that the typical spatial resolution (gas cell size) in the in the host CGM at $z=0$ ranges from $\approx0.06-4\,\kpc$ within the host haloes ($10-500\,\kpc$ from the host).
Whereas, the spatial resolution in low-mass galaxies is usually smaller (better resolved) with cell sizes of $\sim100$ pc, because the gas in the ISM of low-mass galaxies is denser than the gas in the host halo CGM.
Thus, hydrodynamic interactions like ram pressure may be under-resolved far from host galaxies.
However, given that we do not expect especially strong or effective ram pressure stripping far from galaxies, we focus our analysis on ram pressure when galaxies are at small separations from a host or other low-mass galaxies anyway.

\subsection{Ram pressure histories}\label{subsec:rp_histories}

Though observations of galaxies are limited to single snapshots in time of ram pressure effects, we are able to leverage ram pressure histories over the last 10 Gyr of the simulations with $\approx25$ Myr snapshot cadence.
We select all galaxies that are within 2 Mpc of a MW-mass host at $z=0$ and track their trajectories back in time with merger trees.
We calculate the ram pressure they experience as described in Section~\ref{sec:simulations}.
We categorize galaxies by their environment at each snapshot: satellites are within a MW-mass halo\footnote{We use $d_\mathrm{host}(t)<R_\mathrm{200m,\,host}(t)$ as our satellite criterion, which yields 248 satellite galaxies at $z=0$ compared to 240 when using $d_\mathrm{host}(z=0)<300$ kpc because host radii are typically $300-400$ kpc \citep{Samuel2022}.}, galaxies in low-mass groups are within a halo of $\Mtwohm\sim10^{9-11}\,\Msun$ outside of a MW halo, and centrals are not within any halo more massive than their own.
We note that individual galaxies may have belonged to all three groups at different times in the last 10 Gyr, but they are assigned to only a single group per snapshot.

\begin{figure}
    \includegraphics[width=\columnwidth]{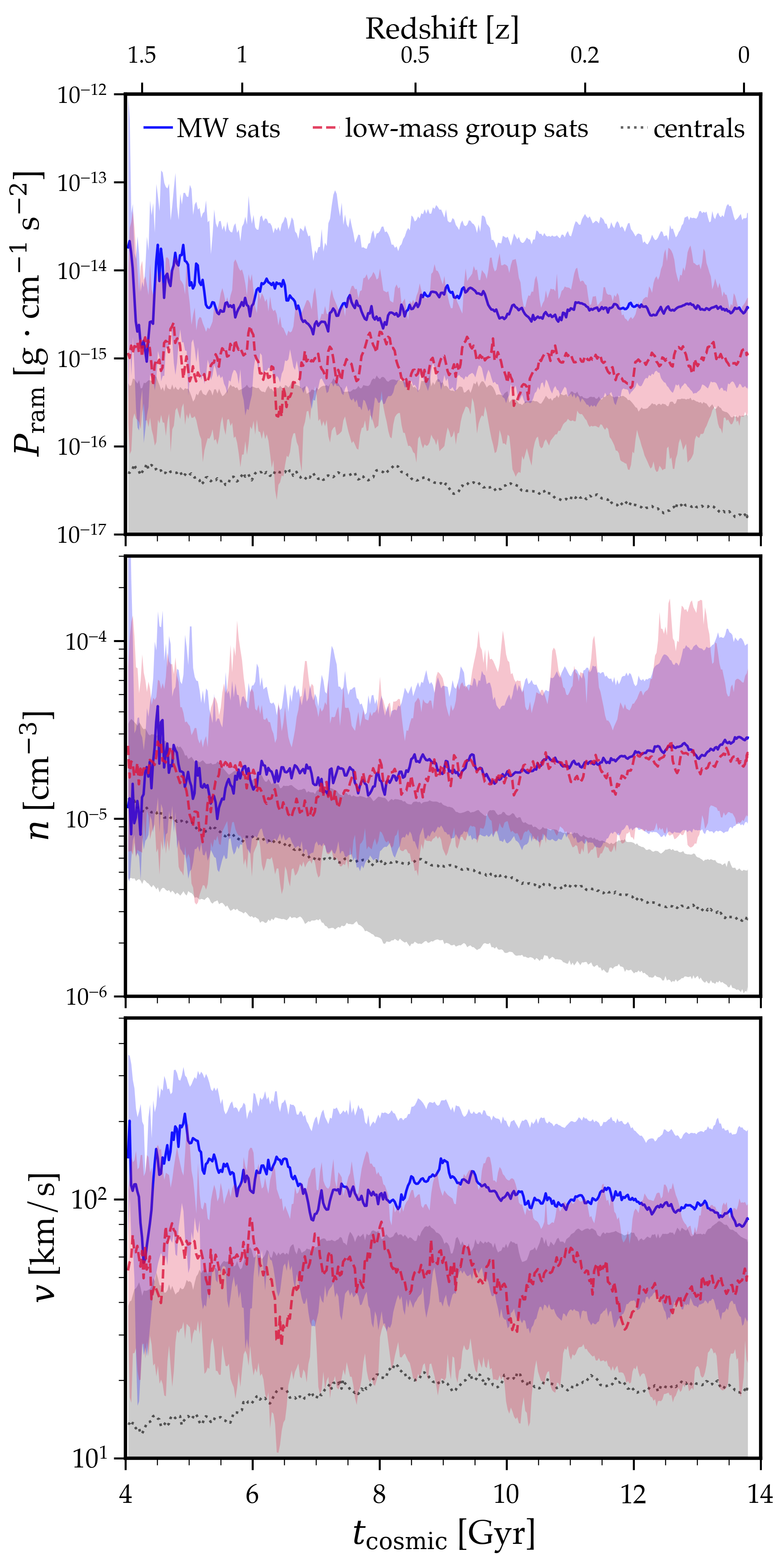}\par
    \vspace{-4 mm}
    \caption{Ram pressure, ambient gas density, and galaxy velocity relative to ambient gas versus time for galaxies in three different environments: MW satellites, galaxies in low-mass groups, and isolated centrals. 
    Lines are medians and shaded regions are 68 per cent variation across all galaxies and simulations at each snapshot.
    \textit{Top:} Ram pressure is highly variable over time within each environment.
    However, across the whole sample, galaxies within an environment experience characteristic, relatively constant levels of ram pressure over time.
    Galaxies in low-mass groups experience a characteristic level of ram pressure that is only about a factor of five lower than that of MW satellites on average.
    \textit{Middle:} The ambient densities encountered by MW satellites and galaxies in low-mass groups are roughly equal, and they gradually increase over time. 
    In contrast, ambient densities encountered by centrals gradually decrease over time.
    \textit{Bottom:} The relative velocities between galaxies and ambient gas drive the overall differences in ram pressure histories between the three environments in the top panel.
    Velocities for MW satellites and galaxies in low-mass groups slightly decrease over time, compensating for the increased ambient densities to yield the relatively constant median ram pressure histories for each environment.
    }
    \label{fig:rp_vs_time}
\end{figure}

Figure~\ref{fig:rp_vs_time} shows the median and 68 per cent variations in ram pressure versus time for low-mass galaxies within these different environments.
As expected, MW satellites experience the highest levels of ram pressure throughout time, as they are in the most dense/crowded environment within our simulation volumes.
Interestingly, ram pressure on MW satellites is only about 5.6 times stronger on average than ram pressure on galaxies in low-mass groups, even though the stellar masses of low-mass group hosts ($M_*\sim10^{7-9}\,\Msun$) are up to 3 dex lower than the MW hosts.
This difference is largely accounted for by different relative velocities between galaxies and ambient gas in the two environments: MW satellites have velocities that are about 2.2 times higher on average, which could account for up to a factor of 4.8 difference in ram pressure from the squared velocity factor in Equation~\ref{eq:local_rp}.
Satellites have the highest average velocity relative to ambient gas at $\sim100$ km/s, followed by galaxies in low-mass groups at $\sim50$ km/s, and centrals have the lowest at $\sim20$ km/s.

However, the median ambient gas densities that galaxies encounter in low-mass groups are roughly equal to those in MW haloes, likely because most galaxies in low-mass groups typically orbit to within $\lesssim50$ kpc from the low-mass host where density is quite high, whereas only half of (surviving) MW satellite galaxies have orbited within 50 kpc of the host \citep{Santistevan2023}.
Centrals, on the other hand, experience relatively low levels of ram pressure because of their comparatively low velocities relative to ambient gas and the lower densities of the ambient gas that they encounter.
Elevated ram pressure during close interactions between low-mass galaxies also helps explain why quenching is effective in low-mass group environments, especially as pre-processing prior to infall into a massive host halo \citep[e.g.,][]{Fujita2004,Bahe2019,Cortese2021,Samuel2022}.

As evidenced by the noisy ram pressure histories, ram pressure is highly stochastic even within a given environment, but in a broad sense the median and scatter for each respective environment has stayed relatively flat over the last 10 Gyr.
The average 68 per cent variation in ram pressure over time on individual low-mass galaxies is roughly 10 times their median ram pressure, regardless of environment.
The galaxy-to-galaxy variation across MW satellites and galaxies low-mass groups at fixed time is also about the same as variation over time for individual galaxies in the respective categories.
Whereas, centrals have about a factor of two higher galaxy-to-galaxy variation at fixed time, largely due to variations in velocity relative to ambient gas.
However, we note that this variability for centrals occurs at ram pressures that are about 1-2 dex lower overall compared to ram pressure on MW satellites or galaxies in low-mass groups.

Interestingly, the ambient gas densities encountered by MW satellites and galaxies in low-mass groups have gradually increased over time, whereas ambient densities encountered by centrals have decreased.
We interpret this as satellites and galaxies in low-mass groups feeling the effects of hierarchical structure formation, whereby their environments have become cluttered with the remnants of past galaxy interactions.
Ambient density near centrals, on the other hand, is probably decreasing because the expansion of the Universe has rendered isolated galaxies farther from gas associated with other galaxies over time.

\begin{figure*}
	\includegraphics[width=\textwidth]{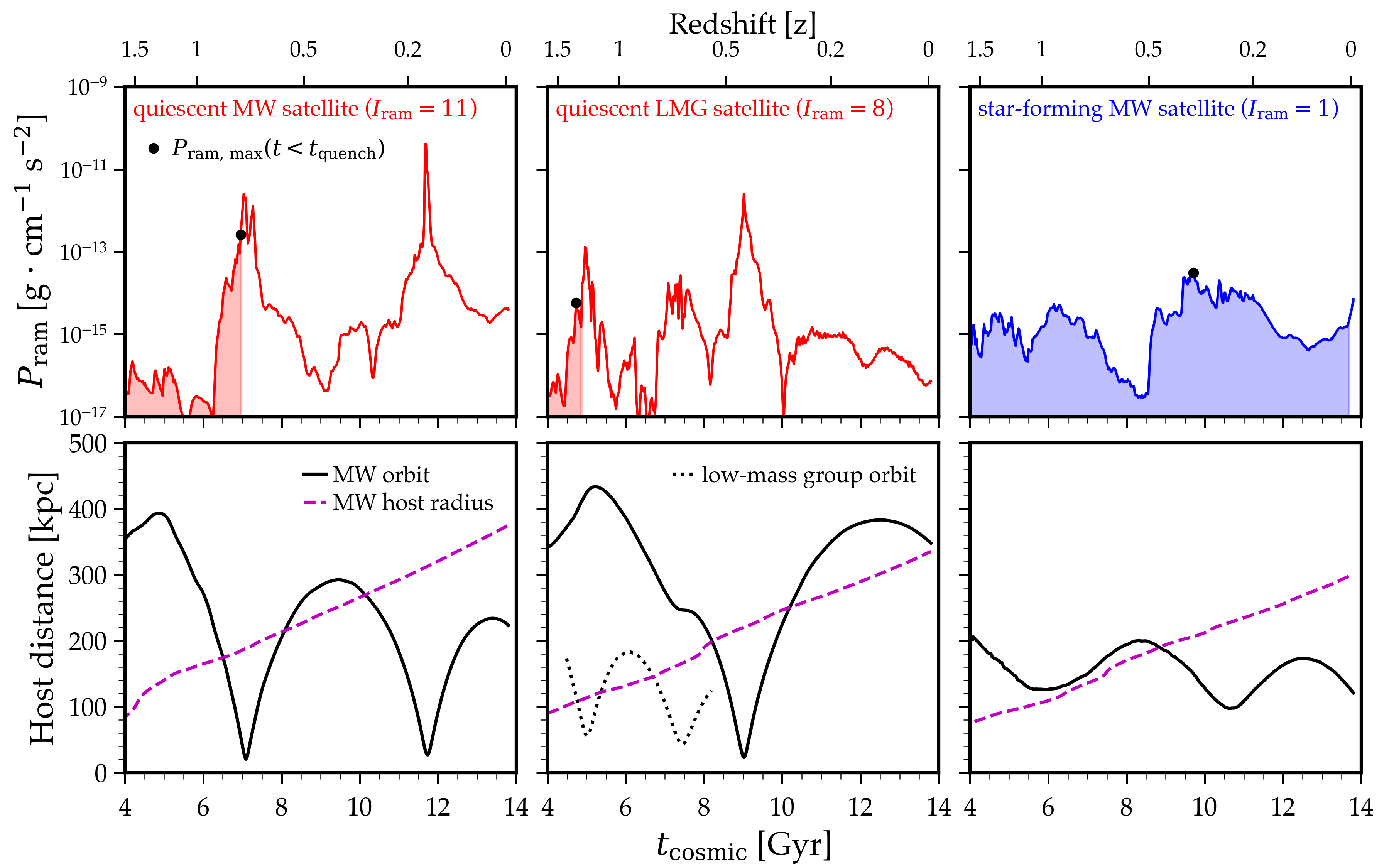}\par
    \vspace{-2 mm}
    \caption{
    Examples of characteristic ram pressure histories (top) and orbits (bottom) for three galaxies that were either MW satellites or were in a low-mass group when they quenched or formed their last star particle.
    The shaded area under each curve indicates the integrated ram pressure felt by that galaxy ($\intPram$), measured from $t=4$ Gyr to when it formed its last star particle, and the black point indicates the maximum ram pressure ($\maxPram$) over that same time period.
    We order the galaxies by decreasing impulsiveness of ram pressure,  $\impulsePram=\maxPram/\intPram$.
    The dashed lines in the bottom panels correspond to the dark matter halo radius ($\rtwoh$) of a MW-mass host, the solid lines are a low-mass galaxy's orbit around the MW-mass host, and the dotted line in the bottom right panel is the low-mass galaxy's orbit around a low-mass group host before it becomes a MW satellite.
    The quiescent galaxies have experienced more impulsive ram pressure compared to the star-forming galaxy, and the peaks in their ram pressure histories tend to coincide with pericentre passages around either a MW host or a low-mass group host.
    }
    \label{fig:example_rp}
\end{figure*}

Figure~\ref{fig:example_rp} shows a few illustrative examples of ram pressure histories for MW satellites and galaxies in low-mass groups.
The ram pressure on the quiescent MW satellite galaxy ($\Mstar=1.3\times10^6\,\Msun$, left panels) varies over about six dex, peaking near pericentre passages that bring it within about 30 kpc of the MW host.
The plunging orbit of the quiescent MW satellite not only brings it near the high density inner regions of the MW halo, but also back out of the halo entirely and into much lower density before splashing back into the MW halo and making a second pericentre passage.
The galaxy in the middle panels of Figure~\ref{fig:example_rp} ($\Mstar=1.2\times10^5\,\Msun$) quenched in a low-mass group before becoming a MW satellite.
Its ram pressure rises by about 3 dex during pericentre passages within the low-mass group, though at an overall lower ram pressure than the quiescent MW satellite.

In contrast, the star-forming MW satellite galaxy (right panels of Figure~\ref{fig:example_rp}, $\Mstar=3.5\times10^7\,\Msun$) has a relatively constant ram pressure history that varies over about two dex.
It experiences a middling increase in ram pressure near infall into the MW halo and first pericentre passage at about 100 kpc from the MW host, and ram pressure rises near $z=0$ as it approaches its second pericentre passage.
We note that the maximum ram pressure these MW satellites experience prior to quenching/last star formation is only $\sim1$ dex higher in the quiescent case than the star-forming case.
Qualitatively, the ram pressure histories of the quiescent galaxies are both strongly peaked compared to the star-forming galaxy, and this is generally true of all galaxies in these samples.
We further quantify the strength of ram pressure peaks relative to the baseline in the following subsection.

\subsection{Ram pressure metrics}\label{subsec:rp_to_quench}

Here we explore the utility of various ram pressure summary statistics that take into account different aspects of the ram pressure histories that we measure.
We explore three key metrics of ram pressure histories in this section: the maximum ram pressure a galaxy experiences prior to quenching ($\maxPram$), an estimate of the ram pressure required to strip gas from and quench a galaxy ($\quenchPram$), and the total integrated ram pressure a galaxy experiences prior to quenching ($\intPram$).

First we compare maximum ram pressure prior to quenching ($\maxPram$) with an estimate of the ram pressure necessary to strip the ISM from a galaxy and thus quench its star formation ($\quenchPram$).
We use a simple estimate for the ram pressure necessary to strip and quench a galaxy, defined as, 

\begin{equation}\label{eq:Pram_to_quench}
    \quenchPram = MAX[2\pi\,\Sigma_\mathrm{dyn}(t) \cdot \Sigma_\mathrm{gas}(t)],
\end{equation}
where $\Sigma_\mathrm{gas}(t)$ is the gas mass surface density within a galaxy over time and $\Sigma_\mathrm{dyn}(t)$ is the combined mass surface density of dark matter and stars within the halo over time (following \citealt{Gunn1972}).
We calculate $\Sigma_\mathrm{gas}(t)$ as described in Section~\ref{sec:simulations}, and we follow a similar methodology for $\Sigma_\mathrm{dyn}(t)$ where we sum the stellar and dark matter masses of a galaxy/halo and divide by the cross-sectional area of the halo assuming spherical symmetry and using the dark matter halo radius, $\rtwoh$.

We note that $\quenchPram$ is an imperfect estimate of the ram pressure required to strip gas from and quench the low-mass galaxies in our simulations.
The equation we use from \citet{Gunn1972} is meant for disc galaxies approaching ambient gas face-on.
In the presence of a true gas disc, the inclination angle between the disc and its motion relative to the ambient gas can change the effectiveness of ram pressure in quenching a galaxy \citep{Bekki2014}.
However, the gas within the low-mass galaxies in our simulations typically has either an irregular or spheroidal morphology because of bursty stellar feedback (see \citealt{Samuel2022} for some visual examples).
Therefore the angle between the major axis of the gas distribution and velocity vector in a low-mass galaxy in our simulations is not necessarily well-defined.

We compute $\quenchPram$ at each snapshot within the last 10 Gyr, and take the maximum value over this time baseline, yielding an approximation of the maximal gravitational restoring force per unit area on the gas in a galaxy over time.
Because of the way we choose to assign gas to a galaxy (using both a distance cut and a velocity cut, see Section~\ref{sec:simulations}) there are frequently no gas cells assigned to a galaxy because stellar feedback increases gas velocity beyond our velocity limits and/or pushes it out of the physical aperture that we use to select the ISM (see Section~\ref{sec:simulations}). 
Using a wider aperture (the subhalo radius) does not change this as it is mainly the velocity cut that excludes gas.
We thus choose to compare $\maxPram$ to $\quenchPram$ rather than taking a maximum of the instantaneous ratio of ram pressure to restoring force per unit area over time, as $\Sigma_\mathrm{gas}$ is frequently zero. 
We can therefore compare $\quenchPram$ to the ram pressure a galaxy experiences to get an idea of whether ram pressure has ever been sufficient to completely strip gas from a galaxy.

\begin{figure*}
\begin{multicols}{2}
\includegraphics[width=\columnwidth]{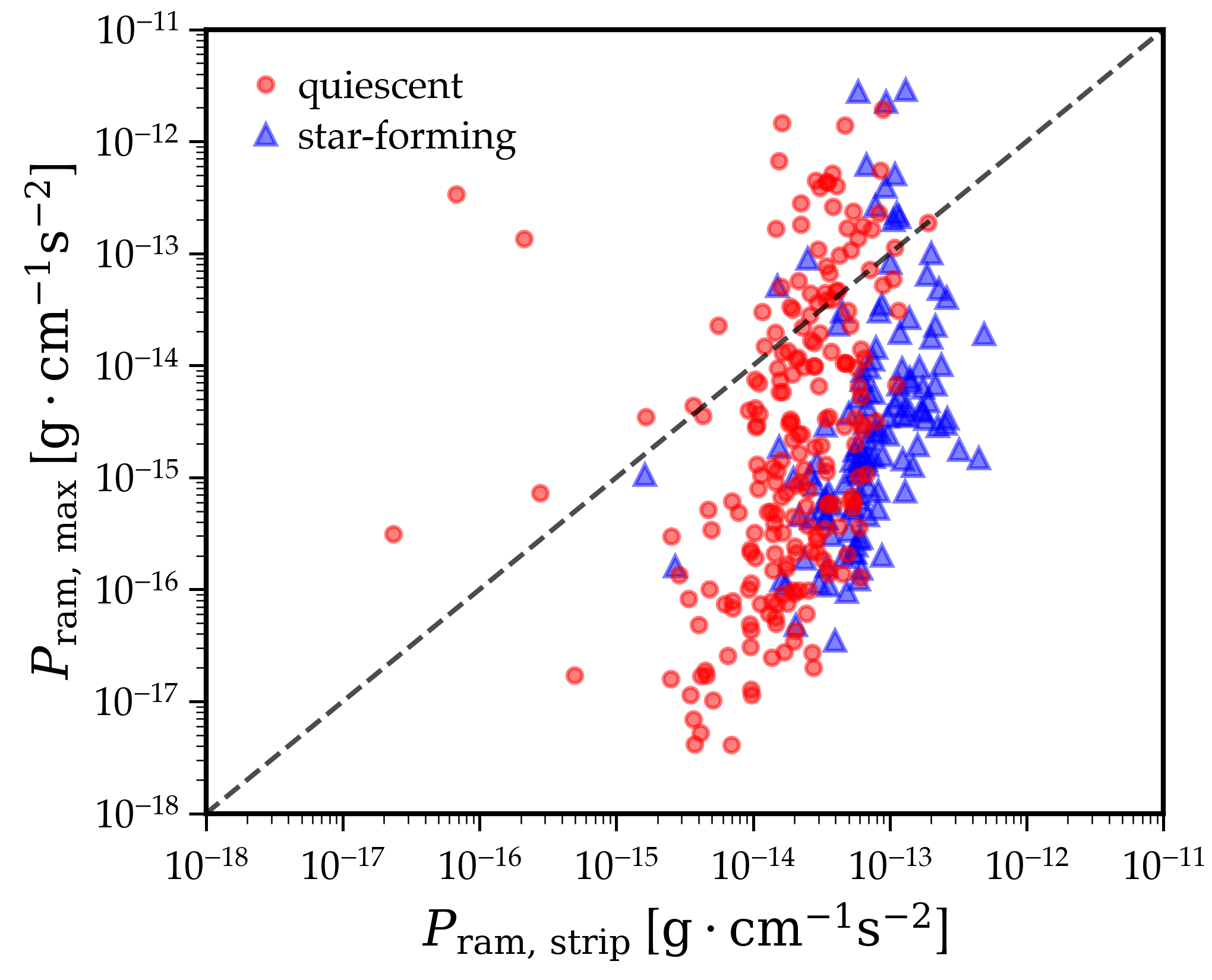}\par
\includegraphics[width=\columnwidth]{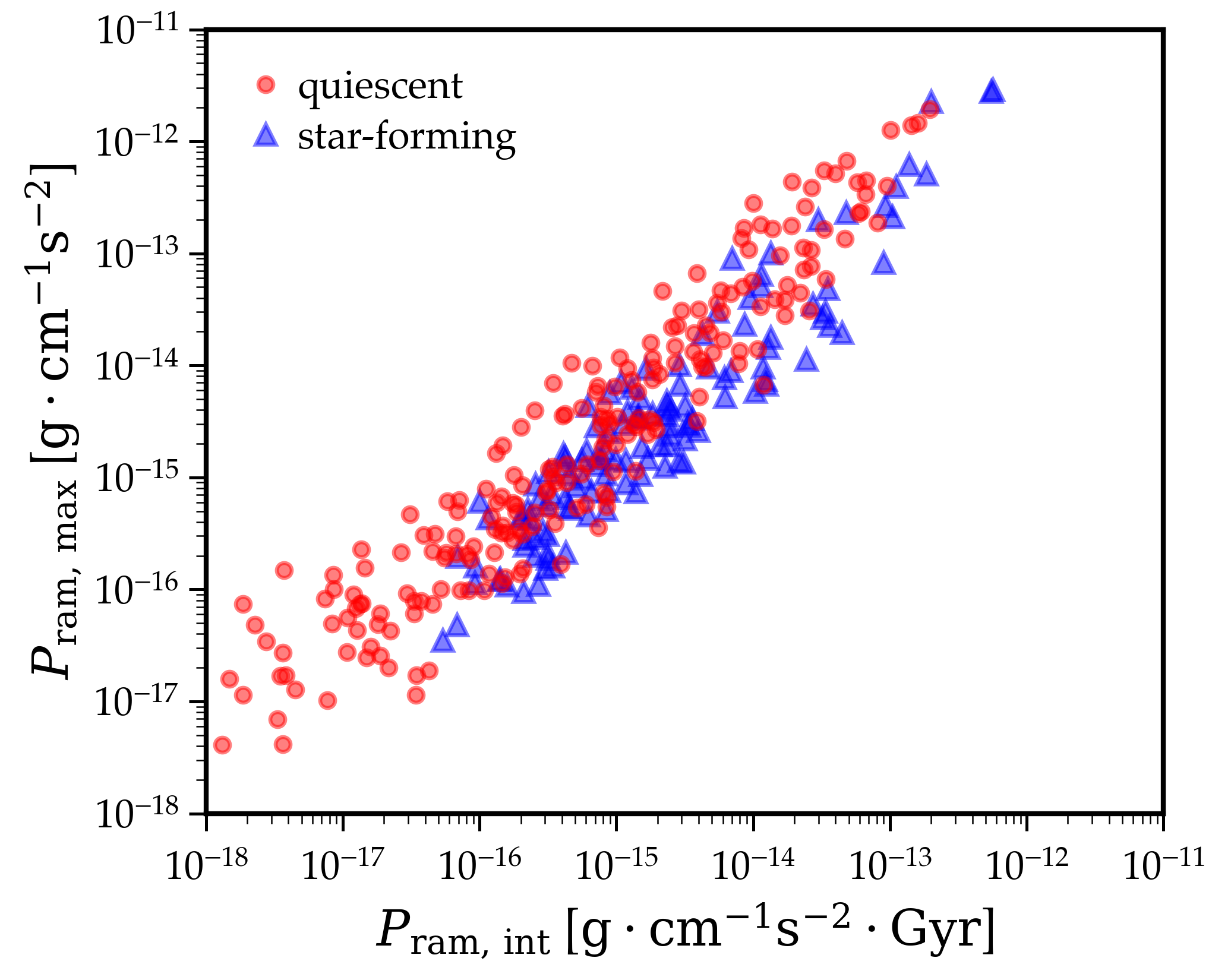}\par
\end{multicols}
    \vspace{-7 mm}
    \caption{
    \textit{Left:} The maximum ram pressure that low-mass galaxies have experienced prior to quenching versus the estimated ram pressure necessary to strip their gas and quench them.
    Quiescent and star-forming galaxies follow a positive trend inclined relative to and blended across the one-to-one line, indicating that ram pressure alone may not have been sufficient to quench galaxies below the one-to-one line, and star-forming galaxies above it have managed to retain gas for star formation despite high levels of ram pressure.
    \textit{Right:} The maximum ram pressure that low-mass galaxies have experienced  versus their integrated ram pressure, prior to when they quenched or formed their last star particle. 
    Quiescent galaxies lie slightly above star-forming galaxies on average, indicating that quiescent and star-forming galaxies have different characteristic average ratios of their maximum to integrated ram pressure.
    }
    \label{fig:rp_estimate}
\end{figure*}

Figure~\ref{fig:rp_estimate} (left) shows the maximum ram pressure experienced by galaxies prior to quenching ($\maxPram$) against our estimate of the ram pressure required to strip their gas and quench them ($\quenchPram$).
Most of the galaxies above the one-to-one line are quiescent at $z=0$, as expected because they have experienced higher ram pressure than necessary to quench.
However, even though almost all of the star-forming galaxies lie below the line, as expected, there is also a significant population of quiescent galaxies below the line.
The quiescent galaxies below the line may indicate the effectiveness of other processes like stellar feedback in quenching galaxies.
The scatter of both quiescent and star-forming galaxies above and below the one-to-one line and the fact that their trends are inclined relative to this line, show that maximum ram pressure and our simple estimate of ram pressure required to quench cannot fully disentangle the quiescent and star-forming populations.

Figure~\ref{fig:example_rp} demonstrates one of the defining features of the ram pressure histories of quiescent galaxies: large peaks above a relatively flat baseline.
We use this fact to motivate our third metric of ram pressure histories, the total integrated ram pressure a galaxy experiences prior to quenching ($\intPram$).
We define integrated ram pressure as,

\begin{equation}\label{eq:Pram_int}
    \intPram = \sum_{i}^{} (P_\mathrm{ram}(t)\cdot\Delta t),
\end{equation}
where the sum is performed over all snapshots \textit{prior to quenching} and the average inter-snapshot spacing is $\Delta t \approx 25$ Myr.
Integrated ram pressure is depicted in Figure~\ref{fig:example_rp} as the shaded regions underneath the ram pressure histories.
In general, integrated ram pressure increases with galaxy stellar mass, from $\sim10^{-17}$ to $\sim10^{-12}$ ${\rm g \cdot cm^{-1} s^{-2} \cdot\,}$Gyr because it takes higher and more prolonged ram pressure to quench more massive galaxies.
We note that integrated ram pressure has units of surface momentum density, and this is how \citet{Simons2020} refer to it.

Figure~\ref{fig:rp_estimate} (right) shows the maximum ram pressure experienced by galaxies against their integrated ram pressure.
The populations of quiescent and star-forming galaxies separate along two parallel but offset relations in this space, with some overlapping scatter.
Star-forming galaxies are concentrated along a tight relationship, and quiescent galaxies typically lie slightly above them with a similar slope.
This indicates that quiescent galaxies feel a strong spike in ram pressure, with a characteristic average ratio of the maximum to integrated values of their ram pressure.
Note that we only include galaxies that contain gas and have not quenched before the 10 Gyr window where we measure ram pressure histories.

\begin{figure*}
\begin{multicols}{2}
	\includegraphics[width=\columnwidth]{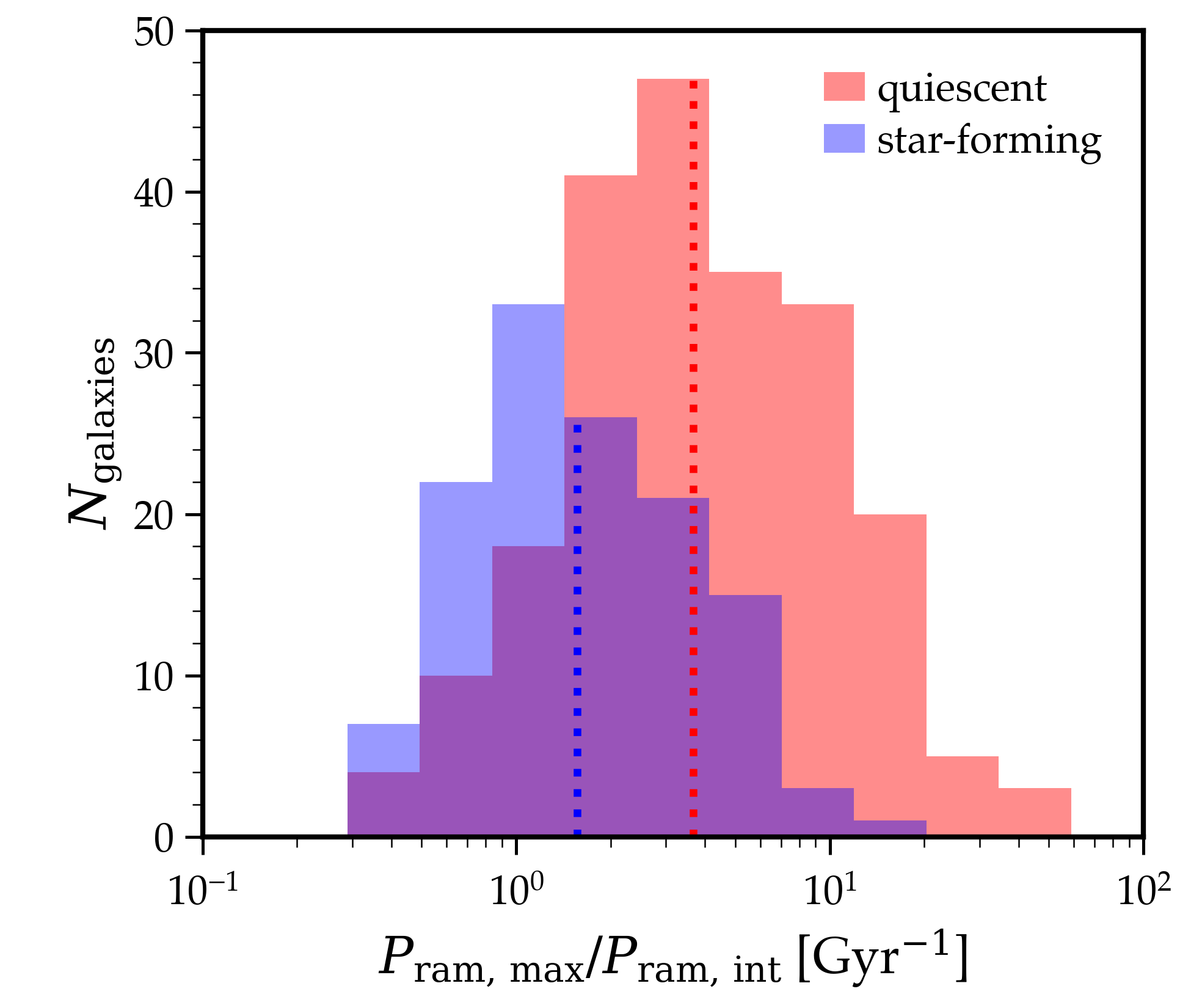}\par
	\includegraphics[width=\columnwidth]{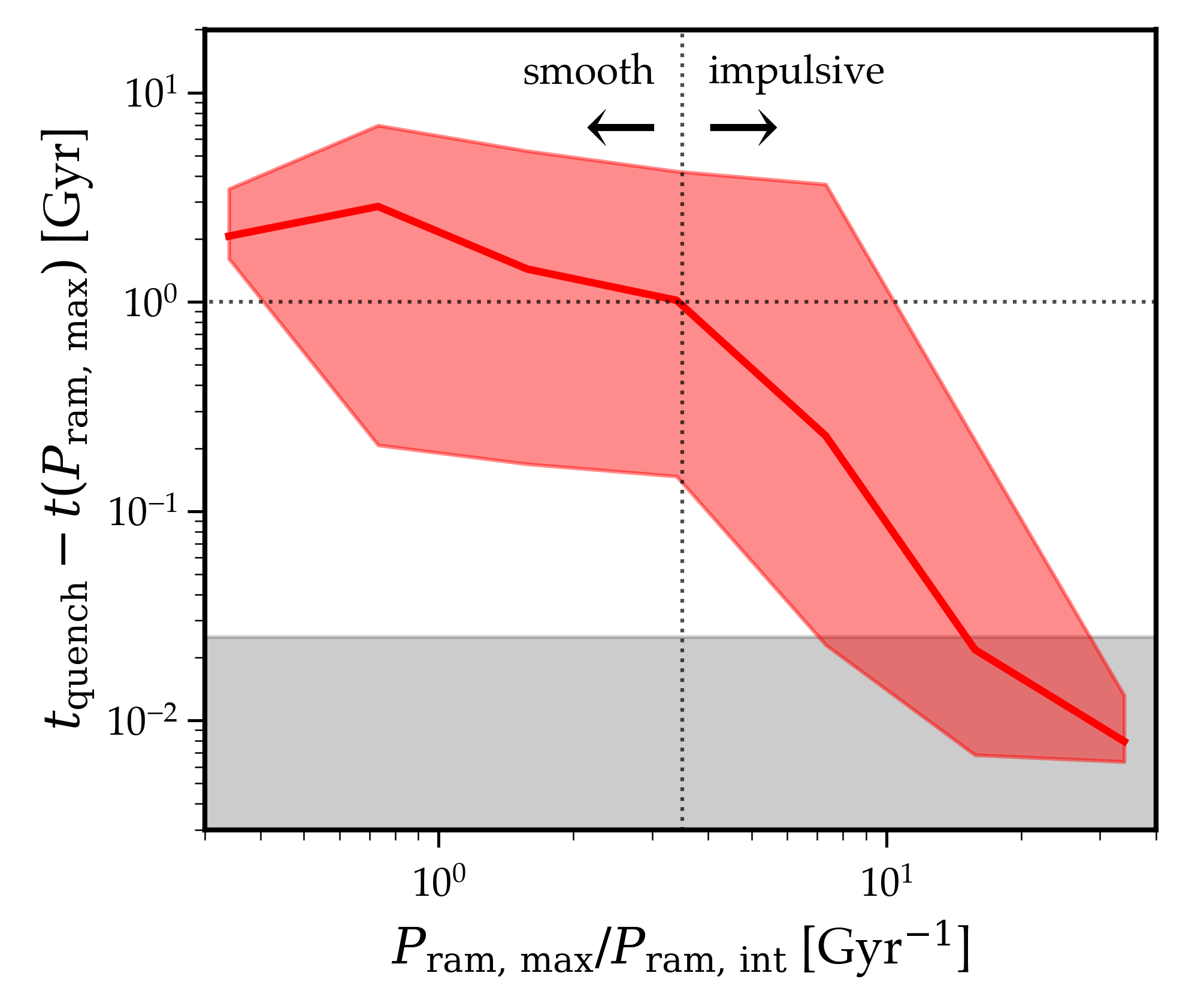}\par
\end{multicols}
    \vspace{-7 mm}
    \caption{
    \textit{Left:} A histogram of the ram pressure impulsiveness ($\impulsePram=\maxPram/\intPram$) for low-mass galaxies. 
    Lines mark the median values of each distribution. 
    Galaxies with $\impulsePram\gtrsim3\,\rm{Gyr^{-1}}$ are highly likely to be quiescent.
    \textit{Right:} The time-scale between maximum ram pressure and quenching as a function of the impulsiveness of ram pressure.
    The line is the median and the red shaded region is the 68 per cent variation among quiescent galaxies.
    The grey shaded region shows the limit of our 25 Myr time resolution/snapshot spacing.
    The typical quenching delay time substantially decreases from about 2 Gyr to $\lesssim100$ Myr as impulsiveness increases.
    Once impulsiveness exceeds $\approx3.5\,\rm{Gyr^{-1}}$, galaxies typically quench within $<1$ Gyr from the time of maximum ram pressure, marking the transition from smooth ram pressure stripping to impulsive ram pressure stripping that rapidly quenches star formation.
    }
    \label{fig:rp_ratio_and_qdt}
\end{figure*}

\subsubsection{Ram pressure impulsiveness}\label{subsec:impulsiveness}

Comparing the left and right panels of Figure~\ref{fig:rp_estimate}, we posit that the defining aspect of a ram pressure history that correlates with whether or not a galaxy is quiescent or star-forming is the \textit{impulsiveness} of ram pressure, which we define as the ratio of maximum ram pressure to integrated ram pressure, 

\begin{equation}\label{eq:rp_impulse}
    I_{\rm ram} = \frac{\maxPram}{\intPram} = \frac{MAX(P_\mathrm{ram}(t))}{\sum_{i}^{} (P_\mathrm{ram}(t)\cdot\Delta t)},
\end{equation}
in units of Gyr$^{-1}$. 
In Figure~\ref{fig:example_rp}, we note that just before quenching, the quiescent galaxies experience a surge in ram pressure of over $2-3$ dex.
Using Equation~\ref{eq:rp_impulse}, we find that the quiescent MW satellite galaxy has $\impulsePram\approx11$ Gyr$^{-1}$, compared to $\impulsePram\approx1$ Gyr$^{-1}$ for the star-forming MW satellite galaxy shown.

We quantify this dichotomy further in Figure~\ref{fig:rp_ratio_and_qdt}.
In the left panel, we show that the distribution for star-forming galaxies peaks at $\impulsePram\approx1\,\rm{Gyr^{-1}}$ while the quiescent distribution peaks at $\impulsePram\approx3\,\rm{Gyr^{-1}}$.
The star-forming distribution has a small tail to higher $\impulsePram$, but in general galaxies with $\impulsePram\gtrsim3$ are highly likely to be quiescent.
About 80 per cent of galaxies with $\impulsePram>3\,\rm{Gyr^{-1}}$ are quiescent, versus about 50 per cent at $\impulsePram<3\,\rm{Gyr^{-1}}$.

In the right panel of Figure~\ref{fig:rp_ratio_and_qdt}, we show the time-scale or delay time between quenching and maximum ram pressure, $t_\mathrm{quench}-t(\maxPram)$, which is positive by definition, versus the impulsiveness of ram pressure for quiescent galaxies.
The time-scale between quenching and maximum ram pressure is clearly anticorrelated with the impulsiveness of ram pressure, and has a Pearson correlation coefficient of $r=-0.52$.
Notably, once ram pressure exceeds an impulsiveness of $\impulsePram\gtrsim3.5\,\rm{Gyr^{-1}}$, the quenching delay time typically drops below 1 Gyr.
Thus, an impulsiveness value of $\impulsePram\approx3.5\,\rm{Gyr^{-1}}$ can be thought of as the division between smooth ram pressure stripping that leads to prolonged quenching or impulsive ram pressure stripping that leads to rapid quenching.
For impulsiveness values $\gtrsim10$, quenching delay times drop to $\lesssim100$ Myr, indicating extremely rapid quenching that approaches the limits of our time resolution.
In Appendix~\ref{sec:metrics}, we show that impulsiveness separates the populations of quiescent and star-forming galaxies the best and has the strongest correlation with quenching time-scale out of five ram pressure metrics that we test.

\subsection{Ram pressure at pericentre}\label{subsec:rp_at_peri}

Because the CGM in the inner host halo is denser than in the outer host halo (see Section~\ref{sec:host_cgm}), and because satellites move fastest in their orbits when they are near pericentre, we expect that satellites should experience stronger ram pressure during pericentre passages.
We identify pericentre passages following \citet{Samuel2022} and \citet{Santistevan2023}, and we specifically examine the pericentre passages that occur closest to when galaxies quench.
Such a pericentre passage may have occurred just before or after quenching in a galaxy, but even in the cases where it has occurred after quenching it may be that the approach to pericentre was the leading cause of quenching.

\begin{figure*}
    \begin{multicols}{2}
	\includegraphics[width=\columnwidth]{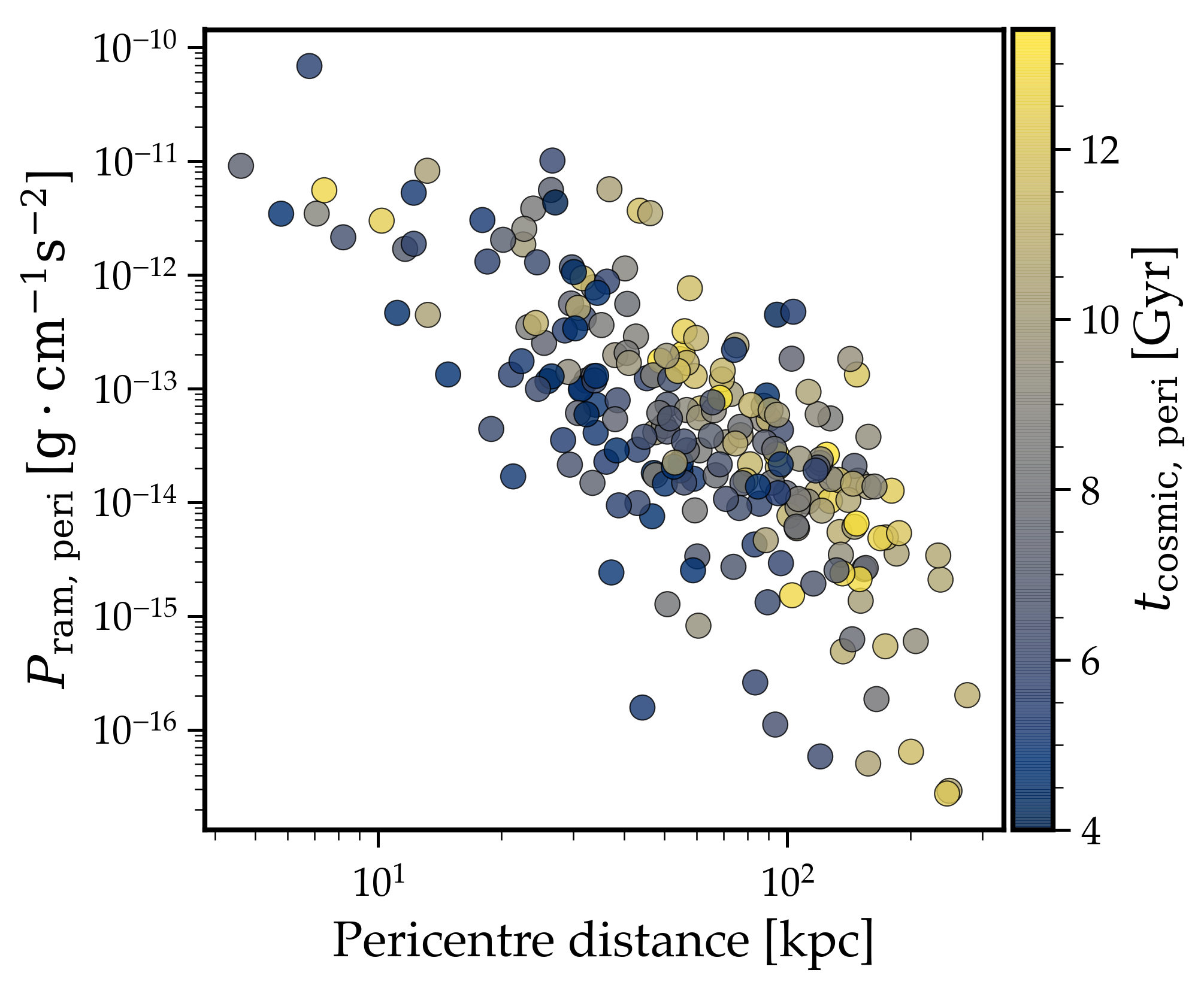}\par
	\includegraphics[width=\columnwidth]{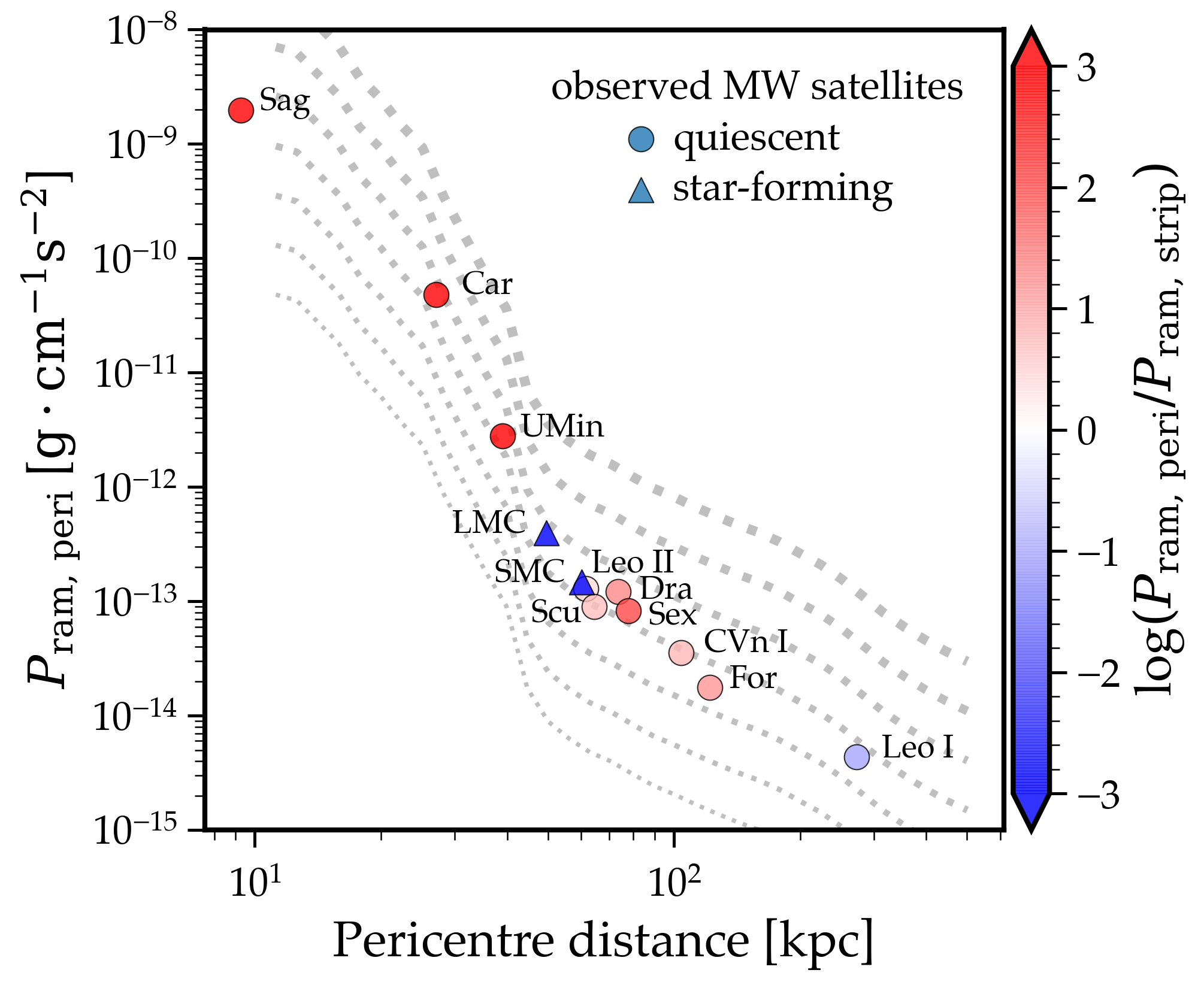}\par
    \end{multicols}
    \vspace{-7 mm}
    \caption{
    \textit{Left:} Ram pressure on quiescent MW satellite galaxies during the pericentre passages that occurred closest to their quenching times. 
    Ram pressure during pericentre passage rises sharply with decreasing distance to the host.
    At fixed physical distance, more recent pericentre passages are typically associated with higher ram pressure, because the median host CGM density in the inner halo increases at late times.
    \textit{Right:} Estimated ram pressure on observed MW satellite galaxies at their most recent pericentre passage. 
    We use the median simulated host CGM density at $z=0$ as a function of distance combined with the distances and velocities of observed satellites at their most recent pericentre passage from integrating their orbits.
    The dotted grey lines are ram pressure profiles at fixed velocities ($50-1000$ km/s), given our median host CGM density.
    Points are coloured by the ratio of ram pressure at pericentre to estimated ram pressure required to quench.
    The estimates using the simulations are similar to observations of MW satellites: most galaxies, except the Magellanic Clouds (blue triangles), are likely to be quiescent at $z=0$ because they have experienced greater ram pressure at pericentre compared to the estimate ram pressure required to quench them.
    Leo I (light blue circle) may have quenched via smooth ram pressure or a localized density spike that is not captured in our median host CGM profile.
    }
    \label{fig:rp_peri}
\end{figure*}

Figure~\ref{fig:rp_peri} (left) shows the effects of inner CGM density enhancement on the ram pressure that satellite galaxies experience during pericentre passage. 
Ram pressure at pericentre ($\periPram$) increases by about four dex as pericentre distance decreases from $\approx100$ kpc to $\approx10$ kpc.
Smaller pericenter distance is also correlated with stronger ram pressure impulsiveness, with a Pearson correlation coefficient of $r=-0.37$ (see previous section for more on impulsiveness).
The 1 dex increase in inner CGM density of some hosts at late times drives a secondary trend (color gradient) in ram pressure at pericentre passage whereby ram pressure is higher for more recent pericentre passages at a fixed physical distance.
We have verified that restricting pericentre passages to those that occur within 2 Gyr of quenching does not change our main results.

Though the median host CGM densities in Figure~\ref{fig:host_cgm} (left) do not accurately represent the clumpy nature of the host CGM, we can still use them to estimate the typical ram pressure felt by observed MW satellite galaxies at pericentre, and weigh this against the ram pressure required to strip their gas and quench them.
To explore this, we integrated the orbits of MW satellites over 10 Gyr using \texttt{galpy} and the host galaxy/halo model \texttt{MWPotential2014} \citep{galpy}.
We stored the distance and velocity at pericentre, and mapped the distance to a host CGM density using the median radial density profile across all of the hosts.
We do not attempt to calculate the impulsiveness of ram pressure on observed MW satellites, as this depends on the localized ambient gas density that they encounter, which is not captured by our median host CGM profiles.

Figure~\ref{fig:rp_peri} (right) shows the estimated ram pressure on 12 MW satellites at pericentre.
We colour the points by the ratio of ram pressure at pericentre passage to the ram pressure needed to quench them.
The shape of the points indicates whether they are actually quiescent ($M_{\rm HI}/M_{\rm dyn}\leq0.1$) or star-forming ($M_{\rm HI}/M_{\rm dyn}>0.1$).
We take values of $M_{\rm HI}$ and $M_{\rm dyn}$ from \citet{Putman2021} and \citet{McConnachie2012}, respectively.
The curves in the background show ram pressure throughout the halo at fixed velocities ($50-1000$ km/s, in log-spaced bins) and based on the median CGM density of our simulated hosts.

By simply scaling the ram pressure at pericentre to the ram pressure required to quench a galaxy, we determine that the star-forming galaxies (the Magellanic Clouds, represented by blue triangles) have not experienced enough ram pressure to quench. 
The remaining galaxies are quiescent by our observational definition and have ratios of $\periPram$ to $\quenchPram$ above unity, which, as we showed in the right panel of Figure~\ref{fig:rp_estimate}, typically corresponds to being quiescent.
The one exception is the quiescent MW satellite galaxy Leo I with a pericentre distance near the edge of the host halo that has not brought it into a dense enough region for ram pressure at pericentre to surpass $\quenchPram$.
However, it may have encountered a dense clump of gas in the host halo or elsewhere that our simple estimate using median host CGM density cannot account for.
This shows that even when neglecting the clumpy nature of the MW CGM, ram pressure at pericentre can account for quenching the observed MW satellite galaxies.


\subsection{Angular anisotropy in the inner host CGM}\label{subsec:cgm_anisotropy}

\begin{figure*}
\begin{multicols}{2}
    \centering
	\includegraphics[width=0.75\columnwidth]{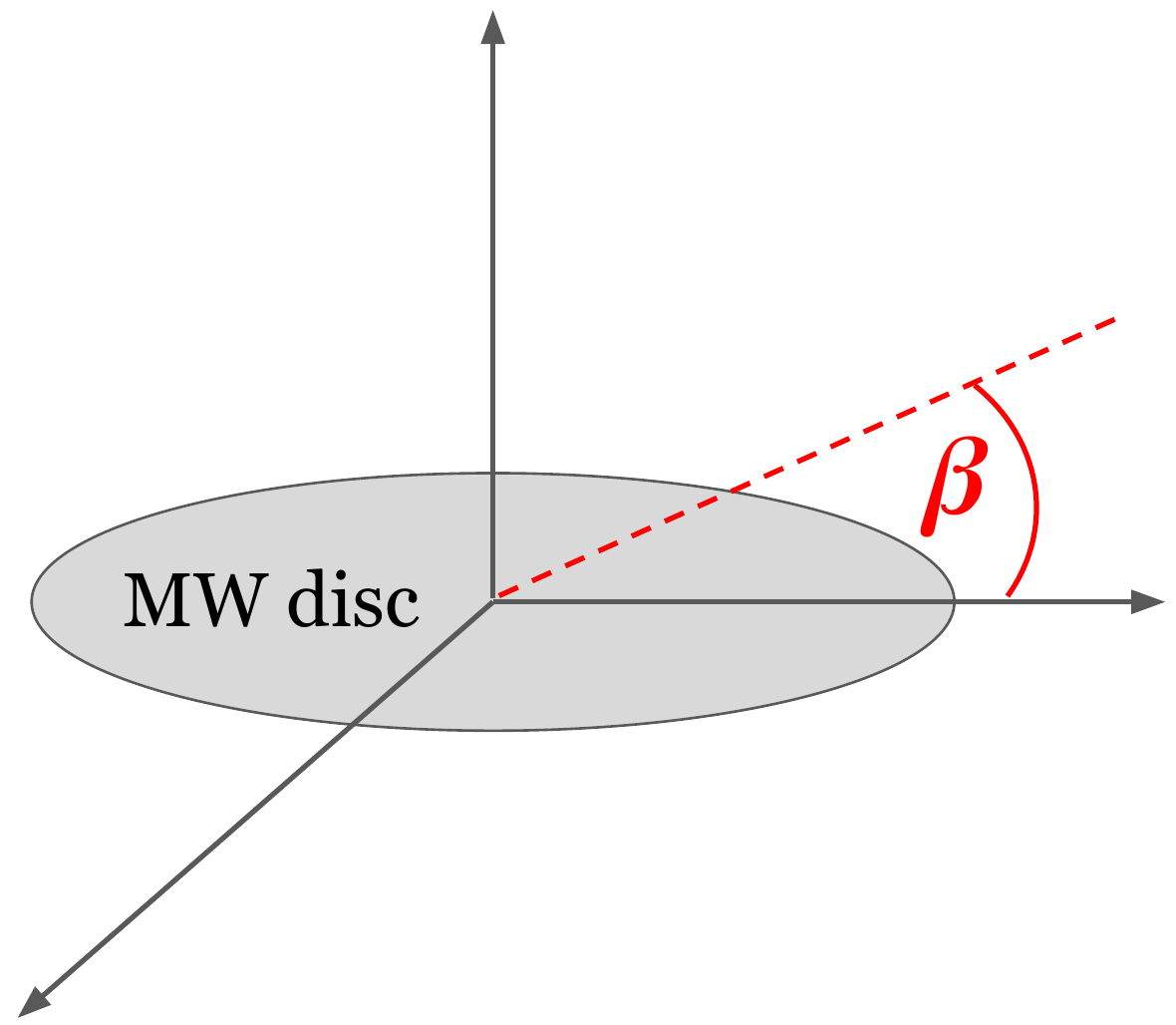}\par
	\includegraphics[width=\columnwidth]{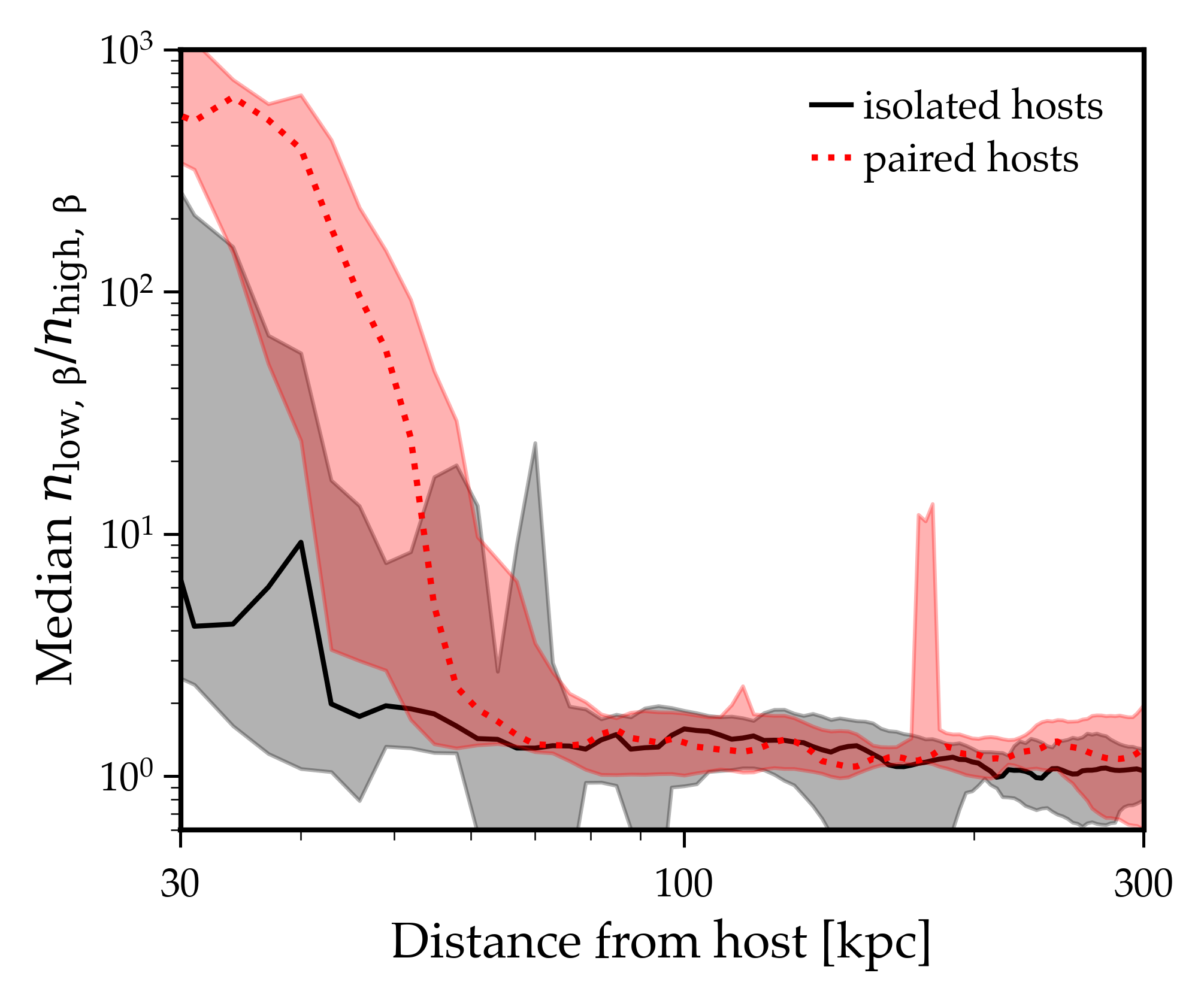}\par
\end{multicols}
    \vspace{-7 mm}
    \caption{\textit{Left:} Schematic of how we measured galactocentric latitude ($\beta$) as the angle from the host galaxy disc. 
    \textit{Right:} The ratio of host CGM density at $z=0$ within low latitudes ($0.8 < |\mathrm{cos}(\beta)| \leq 1.0$) to density at high latitudes ($0 \leq |\mathrm{cos}(\beta)| \leq 0.2$) as a function of distance from the host.
    The lines and shaded regions are the host-to-host medians and 68 per cent variations, respectively. 
    The large spikes near 200 kpc for the paired host distribution and 70 kpc for the isolated hosts are from localized gas overdensities visible near the host disc plane of Louise in the right panels of Figure~\ref{fig:host_vis}, likely from stripped satellite gas.
    The density ratio shows that the inner CGM of both paired and isolated hosts is higher at lower latitudes, and that this effect is stronger for the paired hosts.}
    \label{fig:latitude_ratio}
\end{figure*}

In light of recent work finding anisotropic quenching in the inner regions of massive galaxy clusters \citep[e.g.,][]{MartinNavarro2021}, we search for similar signals close to the MW hosts in our simulations.
We quantify host CGM density, ram pressure, and quenching with respect to galactocentric latitude, 

\begin{equation}\label{eq:latitude}
    \beta = \arctan\left(\frac{z^2}{\sqrt{x^2+y^2}}\right),
\end{equation}
defined as the angle from the host disc, using a Cartesian coordinate system centered on the host halo with $z$ axis perpendicular to the host disc and $x$ and $y$ axes aligned with the host disc.
Note that we measure $\beta$ from the host center, in contrast to the standard galactic latitude ($b$) that is measured from the solar position.
See Figure~\ref{fig:latitude_ratio} (left) for a visual depiction of latitude.
We quantify anisotropy as a function of $|\mathrm{cos}(\beta)|$ in order to sample the host CGM and its satellites uniformly in volume as we vary angle from the host galaxy disc.
This also means that we treat measurements above and below the host disc equally, because we do not have a sufficient number of low-mass galaxies in the inner halo to leverage positive versus negative latitudes.

Figure~\ref{fig:latitude_ratio} (right) shows the ratio of median MW host CGM density in low latitude regions ($0.8 < |\mathrm{cos}(\beta)| \leq 1.0$) to median density in high latitude regions ($0 < |\mathrm{cos}(\beta)| \leq 0.2$) at $z=0$.
We show the host-to-host median and 68 per cent variations, similar to Figure~\ref{fig:host_cgm}.
The median ratios for paired and isolated hosts are relatively flat and near unity at $80-400$ kpc, which indicates that the density in the low and high latitude regions are roughly equal in the outer host halos.
The peaks in the grey scatter around 70 kpc and the red scatter around 200 kpc come from large overdensities near the disc planes of one of the isolated hosts (m12f) and one of the paired hosts (Louise), likely due to recently stripped satellite gas.
However, within $\lesssim60$ kpc the median ratio of density at low versus high latitude increases by about 1 and 2.5 dex for isolated and paired hosts, respectively.
Taken together with our results from Section~\ref{sec:host_cgm}, this means that the inner CGM around paired hosts is anisotropic in density; it is significantly denser within low latitudes at fixed physical distances $\lesssim60$ kpc compared to the isolated hosts.
Given the large physical extent of the gas discs of paired hosts, this could also be viewed as an extension of the disc into the inner halo.

\begin{figure*}
\begin{multicols}{2}
	\includegraphics[width=\columnwidth]{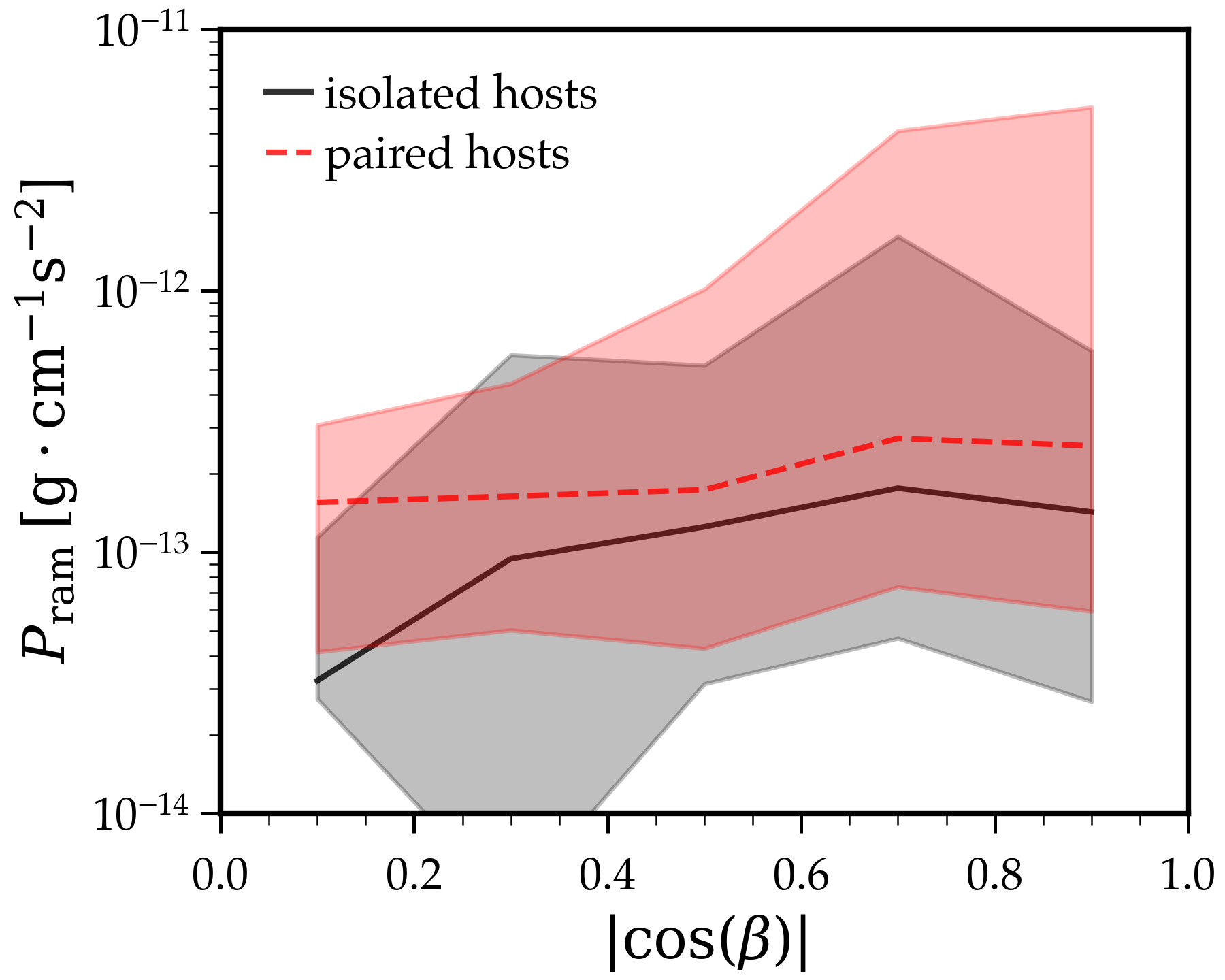}\par
	\includegraphics[width=\columnwidth]{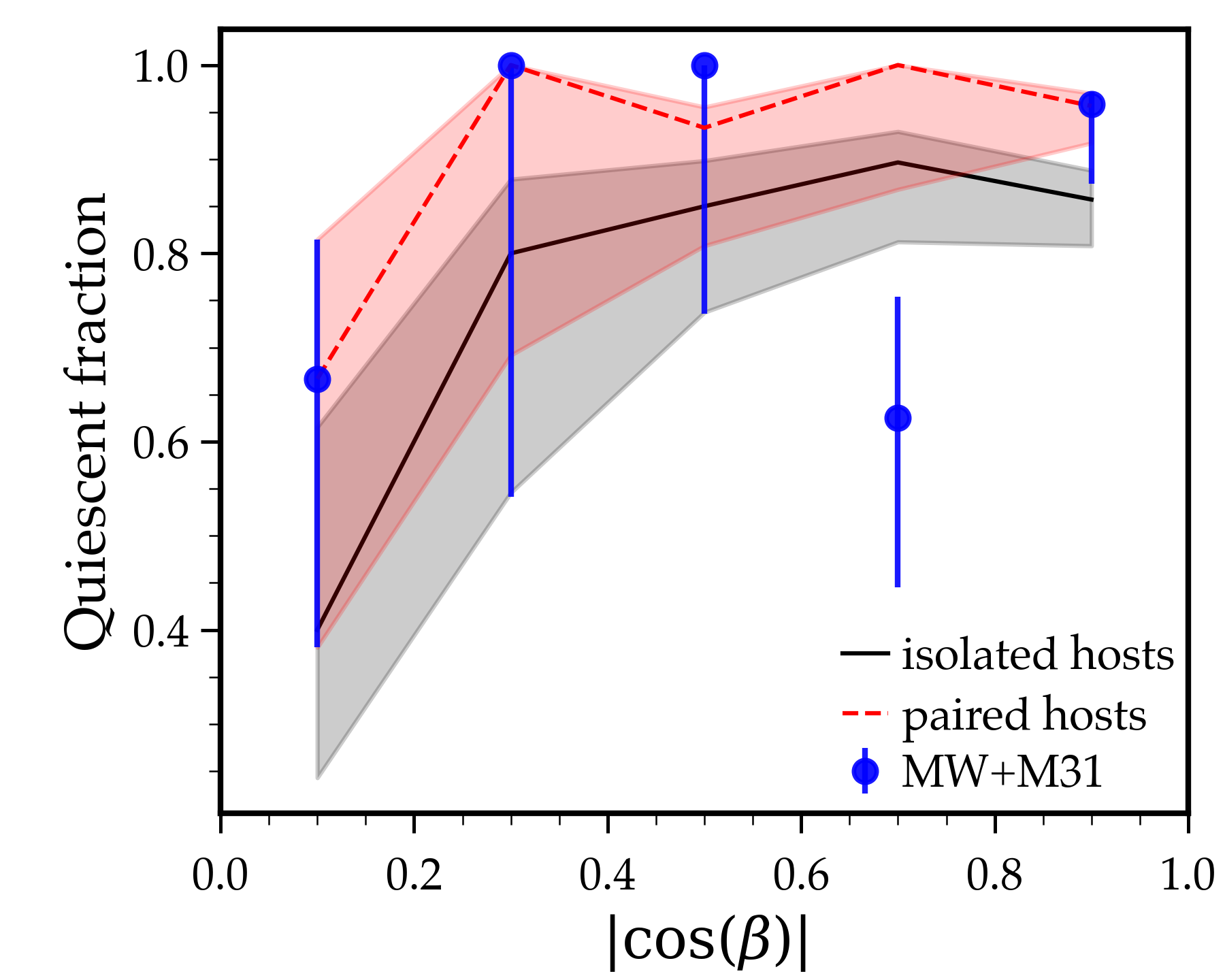}\par
\end{multicols}
    \vspace{-7 mm}
    \caption{\textit{Left:} Ram pressure experienced by surviving satellite galaxies that have orbited within 50 kpc of a host as a function of latitude.
    Lines are medians and shaded regions are 68 per cent variations across hosts, satellites, and time.
    Satellites of paired hosts experience slightly higher median ram pressure at all latitudes compared to satellites of isolated hosts. 
    Ram pressure around paired hosts scatters almost 1 dex higher than isolated hosts at low latitudes ($0.8 < |\mathrm{cos}(\beta)| \leq 1.0$).
    \textit{Right:} The quiescent fraction of satellites at $z=0$ as a function of their latitude.
    The quiescent fraction around simulated hosts rises significantly as latitude decreases.
    Satellites around paired hosts are slightly more quiescent than those around isolated hosts at all latitudes.
    The quiescent fraction of observed satellites around the MW and M31 also generally rises with latitude, and it is mostly consistent with the simulations at the 1-sigma level.
    However, the simulated quiescent fractions are high compared to LG observations at $0.6<|\mathrm{cos}(\beta)|\leq0.8$.}
    \label{fig:latitude_qf}
\end{figure*}

At fixed velocity, higher ambient density should yield a proportionally higher ram pressure.
Figure~\ref{fig:latitude_qf} (left) shows the ram pressure that satellite galaxies have recently experienced within the inner host halo as a function of galactocentric latitude.
We include data for 248 satellite galaxies that have orbited within 50 kpc of a MW-mass host over the last 5 Gyr.
The lines show the medians and the shaded regions show the 68 per cent scatter of all snapshots of satellite orbits meeting these criteria.
The median ram pressure on satellites of paired hosts tends to be slightly greater than that for satellites of isolated hosts at all latitudes.
The ram pressure experienced by satellite galaxies also scatters higher (up to 1 dex) at low latitudes around the paired hosts compared to the isolated hosts.

Higher ram pressures at low latitudes may also correspond to an anisotropic quiescent fraction of satellite galaxies.
Figure~\ref{fig:latitude_qf} (right) shows the quiescent fraction of satellite galaxies versus their angle from the host galaxy disc.
We show 1-sigma uncertainties on quiescent fractions that we calculated using a Bayesian method described in \citet{Samuel2022}.
The quiescent fraction for all hosts shows a distinct rise as latitude decreases (as $|\mathrm{cos}(\beta)|$ increases), indicative of anisotropic quenching. 
The quiescent fraction around paired hosts is also slightly higher than isolated hosts at all latitudes.

We also show the quiescent fraction as a function of latitude for observed LG satellites as points, where we have combined MW and M31 satellites and binned them as a function of $|\mathrm{cos}(\beta)|$ around their respective host.
We use LG satellites in our stellar mass range of $\Mstar=10^{5-10}\,\Msun$ and within 300 kpc of their host, and we consider them to be quiescent if they have $\MHI<10^6\,\Msun$ using data from \citet{Putman2021}.
The quiescent fraction of observed LG galaxies rises similarly to the simulations and is consistent with the simulations at the $\sim1$-sigma level, except for the $0.6 < |\mathrm{cos}(\beta)| \leq 0.8$ bin where the LG has a much lower quiescent fraction compared to the simulations.
We note that the trends with latitude here may be marginal for the LG in part because of small numbers of observed satellites per bin, which are reflected in the large error bars on the points.
If we restrict the satellite stellar mass to $\Mstar>10^6\,\Msun$, we obtain essentially the same results for the observations and simulations.
We interpret the simulation trends in quiescent fraction with latitude as being due to increased ram pressure from the higher density host CGM at low angles from the host disc.
The observed satellites in the LG follow the same general trend as the simulations with the strongest anisotropy in their host CGM, which leads us to conclude that there may be anisotropic host CGM, ram pressure, and satellite quenching present in the LG.

We explored alignments within the paired host simulations that could possibly contribute to such anisotropy, but our findings were inconclusive.
We checked the alignment of each paired host with respect to the vector pointing to the other host galaxy and found values ranging from $2-53\degree$. 
We also checked for alignment between the host galaxy discs in the paired host simulations and found that the angle between the host galaxy disc z axes (normal to the plane of the host disc) varies significantly over the last 10 Gyr, over $\approx20-80\degree$ for all three paired host simulations. 
Given that the anisotropic density that we identify only occurs within $<100$ kpc of each host and the large scatter in the alignments that we measure, it is unclear exactly how alignments between the paired hosts may be related to excess ram pressure or quenching in our analysis.



\section{Discussion}\label{sec:discussion}

Our analysis shows that quiescent galaxies are more likely to have experienced highly impulsive ram pressure just before quenching.
In contrast, smooth ram pressure does not quench galaxies on short time-scales, but rather acts over a prolonged time-scale to remove gas from a galaxy's CGM and prevent the accretion of fresh gas. 
Even MW satellite galaxies may remain star-forming for long periods of time if their ram pressure histories are relatively constant or smooth.
Thus, impulsive ram pressure is likely responsible for effectively stripping the ISM and quenching low-mass galaxies on time-scales as short as $\lesssim100$ Myr.

Our results are in broad agreement with the idea of smooth ram pressure stripping acting on gas in the outskirts or CGM of a galaxy, rather than on the ISM.
For example, \citealt{McCarthy2008} used hydrodynamic simulations to test an analytical model for smooth ram pressure stripping, and found that gas loss ceases when ram pressure and restoring force are roughly equal, which leads to incomplete gas removal even after 10 Gyr.
Furthermore, our results support the findings of work from idealized simulations that showed that comparisons between simulations and observations must take the variation of the ram pressure profile due to a galaxy’s orbit into consideration, as the most profound quenching effects are felt near pericentre where impulsiveness is high for MW satellites \citep{Tonnesen2019}.

It is interesting to compare our results on ram pressure with those from \citet{Simons2020}, who examined ram pressure around six MW-mass hosts from the FOGGIE simulations.
Lagrangian hydrodynamic simulations (like the FIRE simulations we use here) have higher resolution in higher density regions like a galactic disc, which leaves the lower-density CGM comparatively under-resolved. 
However, FOGGIE simulates the CGM around MW-mass galaxies at uniformly high resolution using the grid-based code Enzo and a forced refinement scheme that pre-tracks the halo of interest in a lower resolution run.  
The MW-mass host haloes of the FOGGIE simulations are resolved with cell sizes of $\approx100-200$ pc, and the resolution sharply degrades to $\approx3-5$ kpc beyond $2R_{vir}\sim700$ kpc \citep{Peeples2019}.
In comparison, our simulations achieve similar resolution in the host haloes on average: our resolution actually surpasses FOGGIE's in the innermost regions ($\dhost<20$ kpc) at $\approx60$ pc, our resolution is roughly equal to FOGGIE's (200 pc) at $\dhost\approx30-40$ kpc, and our resolution gradually increases to about 4 kpc at $\dhost\approx500$ kpc.

Though \citet{Simons2020} concentrate their analysis on $z\geq2$ (prior to when we measure ram pressure histories) because some of their simulations were not run all the way to $z=0$, we discuss in more detail below the aspects of their work concerned with the satellite ram pressure histories that they measure over $\sim0.1-1$ Gyr. 
Our results are qualitatively similar to theirs, as they find high stochasticity in ram pressure on individual galaxies and throughout the host haloes.
Peaks in ram pressure on FOGGIE satellites occur over very short time-scales ($\lesssim$50 Myr), which they are able to resolve given their 5 Myr snapshot spacing, but we are unable to resolve with our $\approx$25 Myr snapshot spacing.

\citet{Simons2020} also explored integrated ram pressure, referred to as surface momentum imparted by ram pressure on a galaxy in their work.
Notably,  \citet{Simons2020} found that galaxies with higher integrated ram pressure (up to $z=2$) tend to have lost a greater fraction of their gas.
While this is also true for quiescent galaxies in our simulations, neither the gas loss from star-forming galaxies nor the overall quiescent fraction of galaxies in our simulations show a significant trend with integrated ram pressure.
This difference could be due to the much shorter time ($\sim0.1-1$ Gyr) over which integrated ram pressure is measured and the smaller snapshot time spacing in FOGGIE leading to better correlation with short term gas loss, as compared to our 10 Gyr time baseline which encompasses many cycles of star formation, stellar feedback, and pericentric passages for satellite galaxies that may wash out such a correlation.

In addition, our results identifying anisotropic quenching, ram pressure, and host CGM are qualitatively similar to a few recent studies that have found signals of anisotropic quenching within both simulated and observed massive galaxy clusters.
\citet{MartinNavarro2021} first reported a statistically significant enhancement of the quiescent fraction of galaxies near the major axis of the brightest cluster galaxies in SDSS ($\Mhalo\sim10^{12-14}\,\Msun$ and median $z=0.08$).
Based on their analysis of a similar signal in the TNG simulations, they concluded that feedback from active galactic nuclei (AGN) drives anisotropy in the host CGM density which in turn drives anisotropy in ram pressure and quenching.

Following \citet{MartinNavarro2021}, other authors have also found anisotropic quenching in massive clusters with the Sloane Digital Sky Survey (SDSS) and at higher redshift ($z\lesssim1.25$) using the Cluster Lensing And Supernova survey with Hubble (CLASH) and the Hyper Suprime-Cam Subaru Strategic Program \citep{Zhang2022,Stott2022,Ando2023}.
Interestingly, \citet{Stott2022} proposed an alternative explanation for anisotropic ram pressure and quenching that does not rely on AGN feedback.
This author suggests that elongation of the CGM and intracluster medium out to large distances around a BCG, similar to the ellipsoidal shape of the BCG itself, could also produce higher ram pressure at low angles from the BCG major axis.
Given the lack of both AGN feedback and correlation between anisotropy and alignment between paired hosts in our simulations, the alternative explanation put forth by \citet{Stott2022} seems sufficient to explain both the enhanced CGM density, quiescent fraction, and ram pressure at low angles from host discs that we find here.
We posit that perhaps the anisotropic deposition of satellite gas or accretion of satellites onto the host may be a possible source of the anisotropy in the CGM of paired hosts.

\subsection{Caveats}

A caveat to our work is that we only examine the ram pressure histories of galaxies that survive to $z=0$.
We may be missing galaxies that experience impulsive ram pressure stripping but are also gravitationally disrupted and/or not identified by the halo finder, so-called `orphan' galaxies.
For example, in our analysis of anisotropy in the host CGM (Section~\ref{subsec:cgm_anisotropy}), we have examined the ram pressure histories of surviving, close-orbiting satellites ($d_{\rm host}<50$ kpc).
If we included galaxies that disrupted before $z=0$ in our analysis, we could increase the number of ram pressure histories that we examine.
Orphan galaxies likely experienced broadly similar ram pressure histories to the MW satellite galaxy population that we study, perhaps analogous to ram pressure histories of surviving satellites with close pericenter passages.
However, the exclusion of orphan galaxies is unlikely to change our results on ram pressure prior to quenching, such as the correlation between ram pressure impulsiveness and rapid quenching, because we measure ram pressure for this part of our analysis while galaxies are still star-forming and typically well-resolved.

Previous work shows that the dynamical evolution of subhaloes hosting low-mass galaxies in our simulations is likely resolved \citep{Samuel2020}, and we have demonstrated that the hydrodynamic spatial resolution within the ISM of low-mass galaxies is comparable to that in the inner host CGM ($\lesssim100$ kpc) where ram pressure is strongest (Appendix~\ref{sec:resolution}).
However, the hydrodynamic spatial resolution in our cosmological simulations is still limited compared to idealized simulations, and thus the correlation that we find between shorter quenching timescales and highly impulsive ram pressure stripping should be vetted against higher resolution simulations.
To this end, we briefly compare ram pressure in our simulations to ram pressure in an ultra-high-resolution (UHR) version of one of our hosts (m12i) simulated to $z=0$ (Wetzel et al., in prep.).
The UHR simulation has $8 \times$ smaller baryonic particle mass (m$_{\rm baryon,ini,UHR}=880\,\Msun$), and about the same minimum hydrodynamic spatial resolution as the simulations used in this work ($\sim1$ pc).
In general, the ram pressure experienced by low-mass galaxies in the UHR simulation is similar to the simulations used in this work, and the quiescent galaxies follow a similar trend in quenching timescale versus impulsiveness as that shown in Figure~\ref{fig:rp_ratio_and_qdt}.

We also have not explicitly examined the role of stellar feedback as an alternative or complementary mechanism to ram pressure in quenching low-mass galaxies.
In particular, supernovae can rarefy and eject a galaxy's ISM, which may then cause it to be more easily removed by ram pressure \citep{ElBadry2016}.
This extended gas may be re-accreted onto isolated galaxies, but if a supernova explosion occurs within a dense host halo environment the ejected gas may be more easily removed by ram pressure.
Supernovae are individually time-resolved in our simulations, but we do not save the explosion information for individual star particles.
Though it would be possible to estimate the timing of supernovae given our underlying stellar model in the simulation in order to correlate stellar feedback with gas removal, we do not pursue such an analysis here.

Moreover, we have not yet explored the detailed and well-studied role of ram pressure \textit{inducing} star formation in galaxies \citep[e.g.,][]{Tonnesen2009,Genina2019,Wright2019,Hausammann2019,DiCintio2021}.
Most of these studies have focused on satellites of massive hosts like the MW, but the connection between ram pressure in low-mass groups and induced star formation remains to be explored.
In particular, \citealt{Massana2022} recently showed that star formation in the LMC and SMC was correlated over the last $\sim3.5\,\gyr$, which may have been induced by the mutual interactions between this galaxy pair \citep{Moreno2019,Moreno2021,Moreno2022}. 
In future work we will look for such correlations in simulated low-mass groups outside of a MW halo or between satellite-satellite pairs within a MW halo.

\section{Conclusions}\label{sec:conclusions}

We have examined the ram pressure experienced by low-mass galaxies in simulations of Local Group-like environments.
We find several trends relating ram pressure and quenching to different aspects of a low-mass galaxy's evolution such as its environment, the impulsiveness of ram pressure it experiences, pericentre passage, and presence in the halo of a LG-like paired host versus an isolated MW host.
Below, we list our conclusions.

\begin{enumerate}
    \item Ram pressure on MW satellite galaxies is only 5.6 times higher on average than ram pressure on galaxies in low-mass groups, despite up to three dex differences in host stellar mass. 
    This helps explain why group pre-processing is an effective quenching mechanism for low-mass LG satellites. 
    \item Quiescent galaxies have experienced more impulsive ram pressure on average. 
    We quantify the impulsiveness of ram pressure ($\impulsePram$) by scaling the maximum ram pressure to the integrated ram pressure (Equation~\ref{eq:rp_impulse}). 
    We find that galaxies with $\impulsePram\gtrsim3\,\gyr^{-1}$ are highly likely to be quiescent, whereas galaxies with lower $\impulsePram$ are equally likely to be quiescent or star-forming. 
    Thus, ram pressure may need to be significantly impulsive to quench a galaxy.
    \item The time-scale between maximum ram pressure and quenching strongly correlates with the impulsiveness of ram pressure ($\impulsePram$). At $\impulsePram\gtrsim3.5\,\gyr^{-1}$, the median quenching delay time decreases to $<1\,\gyr$ and decreases further to $\lesssim100$ Myr at $\impulsePram\gtrsim10\,\gyr^{-1}$. Therefore, the impulsiveness of ram pressure may also dictate how rapidly quenching proceeds.
    \item Ram pressure rises sharply with decreasing distance to the host, by about four dex going from 100 to 10 kpc from the host. At fixed distance, more recent pericentre passages are also typically associated with higher ram pressure because the host CGM is denser at small physical distances at late times.
    \item The host CGM density is larger in the inner halo compared to the outer halo by $2-4$ dex on average across our sample of 14 hosts. In the inner host halo, the paired hosts have typical densities up to two dex above that of the isolated hosts.
    \item The CGM density around paired hosts also varies as a function of angle from the host disc, $\beta$ or latitude, whereby density is enhanced by $\approx2$ dex at low latitudes ($0.8 < |\mathrm{cos}(\beta)| \leq 1.0$) and small distances ($\lesssim60$ kpc) from the disc compared to high latitudes. 
    The CGM around isolated hosts shows a less significant and noiser density enhancement of $\lesssim1$ dex at low latitudes. 
    The median ram pressure on close-orbiting ($<50$ kpc) satellites around paired hosts is higher than for satellites of isolated hosts.
    The anisotropic density and ram pressure enhancements around simulated hosts are also reflected in the anistropic quiescent fraction of ($z=0$) their satellite galaxies that is higher at lower latitudes.
    A similar trend for observed MW and M31 satellite galaxies may indicate anisotropic quenching in the LG as well.
\end{enumerate}

\section*{Acknowledgements}

We thank the referee for their useful comments and suggestions that improved the quality of this paper.
JS was supported by an NSF Astronomy and Astrophysics Postdoctoral Fellowship under award AST-2102729.
BP received support from the REU program at UC Davis through NSF grant PHY-1852581.
JS and AW received support from: NSF grants CAREER 2045928 and 2107772; NASA ATP grants 80NSSC18K1097 and 80NSSC20K0513; HST grants AR-15057, AR-15809, GO-15902 from the Space Telescope Science Institute (STScI), which is operated by the Association of Universities for Research in Astronomy, Inc., for NASA, under contract NAS5-26555; a Scialog Award from the Heising-Simons Foundation; and a Hellman Fellowship.
IS received support from NASA, through FINESST grant 80NSSC21K1845.
MBK acknowledges support from NSF CAREER award AST-1752913, NSF grants AST-1910346 and AST-2108962, NASA grant 80NSSC22K0827, and HST-AR-15809, HST-GO-15658, HST-GO-15901, HST-GO-15902, HST-AR-16159, and HST-GO-16226 from the Space Telescope Science Institute, which is operated by AURA, Inc., under NASA contract NAS5-26555.
JM is funded by the Hirsch Foundation.
CAFG was supported by NSF through grants AST-1715216, AST-2108230,  and CAREER award AST-1652522; by NASA through grants 17-ATP17-0067 and 21-ATP21-0036; by STScI through grants HST-AR-16124.001-A and HST-GO-16730.016-A; by CXO through grant TM2-23005X; and by the Research Corporation for Science Advancement through a Cottrell Scholar Award.

We performed this work in part at the Aspen Center for Physics, supported by NSF grant PHY-1607611, and at the KITP, supported NSF grant PHY-1748958.

We ran simulations using the Extreme Science and Engineering Discovery Environment (XSEDE), supported by NSF grant ACI-1548562; Frontera allocations AST21010 and AST20016, supported by the NSF and TACC; Blue Waters via allocation PRAC NSF.1713353 supported by the NSF; the NASA HEC Program through the NAS Division at Ames Research Center.

We gratefully acknowledge use of the IPython package \citep{ipython}, NumPy \citep{numpy}, SciPy \citep{scipy}, Numba \citep{numba}, matplotlib \citep{matplotlib}, and \texttt{galpy} \citep{galpy}.


\section*{Data Availability}

The FIRE-2 simulations are publicly available \citep{Wetzel2023} at \url{http://flathub.flatironinstitute.org/fire}.
Additional FIRE simulation data is available at \url{https://fire.northwestern.edu/data}.
A public version of the \textsc{Gizmo} code is available at \url{http://www.tapir.caltech.edu/~phopkins/Site/GIZMO.html}.
The publicly available software packages used to analyze these data are available at: \url{https://bitbucket.org/awetzel/gizmo\_analysis}, \url{https://bitbucket.org/awetzel/halo\_analysis}, and \url{https://bitbucket.org/awetzel/utilities} \citep{WetzelGizmoAnalysis2020,WetzelHaloAnalysis2020}.



\bibliographystyle{mnras}
\bibliography{ref}



\appendix

\section{Spatial resolution in the ISM and CGM}\label{sec:resolution}

\begin{figure}
    \includegraphics[width=\columnwidth]{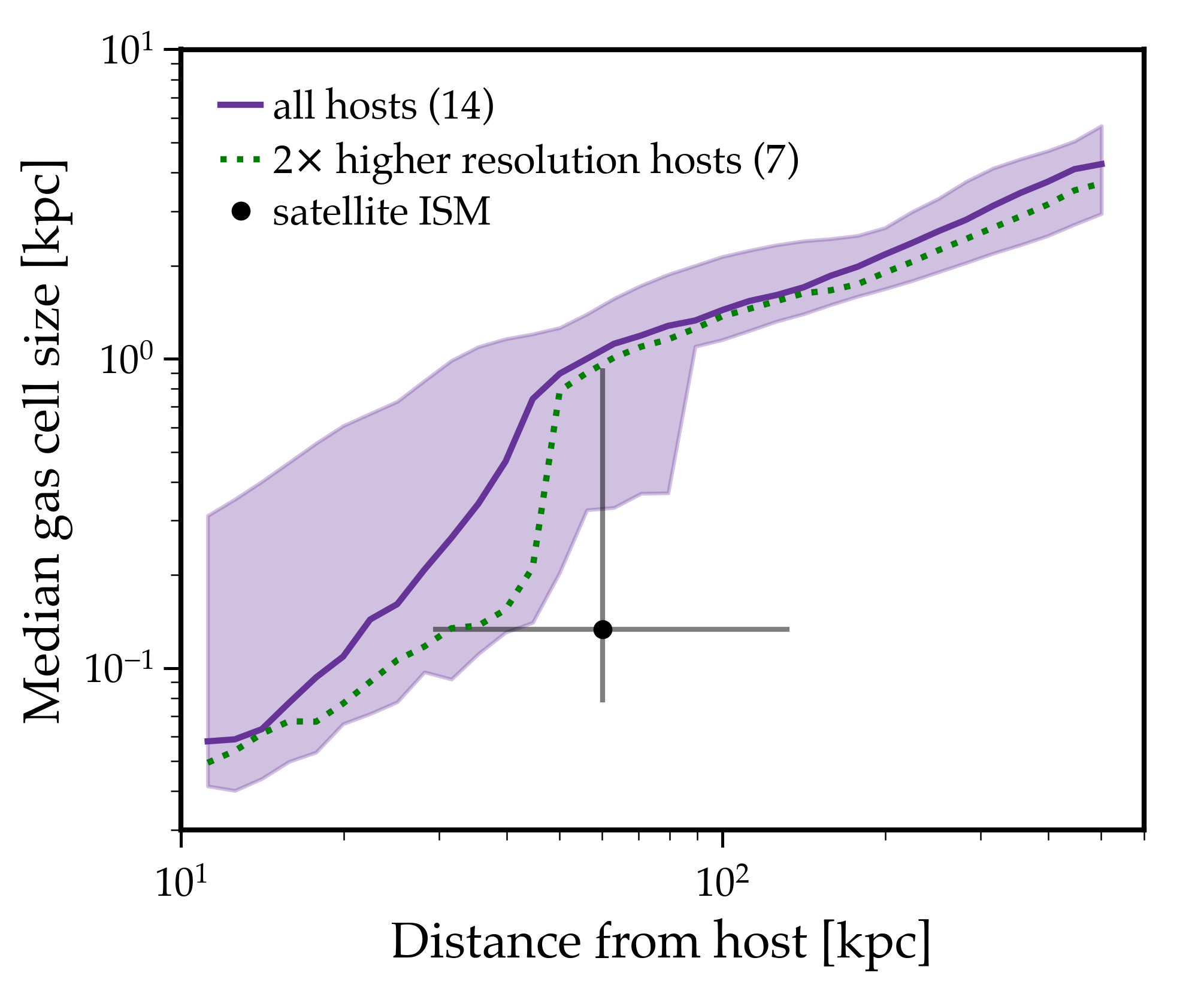}\par
    \vspace{-2 mm}
    \caption{
    The median spatial resolution (gas cell size) in the host CGM at $z=0$ ranges from $\approx0.06-4\,\kpc$ within the host haloes ($10-500\,\kpc$ from the host).
    Shaded regions are the 100 per cent host-to-host variations in their median gas cell sizes, such that the full range of resolution is much larger.
    We also show the typical gas cell size in the ISM of satellite galaxies as a point at the median pericentre distance of all satellites, and the lines extending from it are the 68 per cent ranges across all satellites over time.
    }
    \label{fig:host_cgm_res}
\end{figure}

In Figure~\ref{fig:host_cgm_res}, we show the typical gas cell size in the host CGM versus distance from the host, and the typical cell size in the ISM of satellite galaxies.
Cell sizes in the host halo range from $\lesssim100$ pc in the inner halo to $\sim4$ kpc in the outer halo.
The higher resolution hosts have denser/better resolved gas out to about 50 kpc compared to the fiducial resolution hosts.
Cell sizes in the ISM of low-mass galaxies are typically $\sim100$ pc, comparable to the inner host halo.

\section{Ram pressure metrics}\label{sec:metrics}

\begin{figure*}
\includegraphics[width=\textwidth]{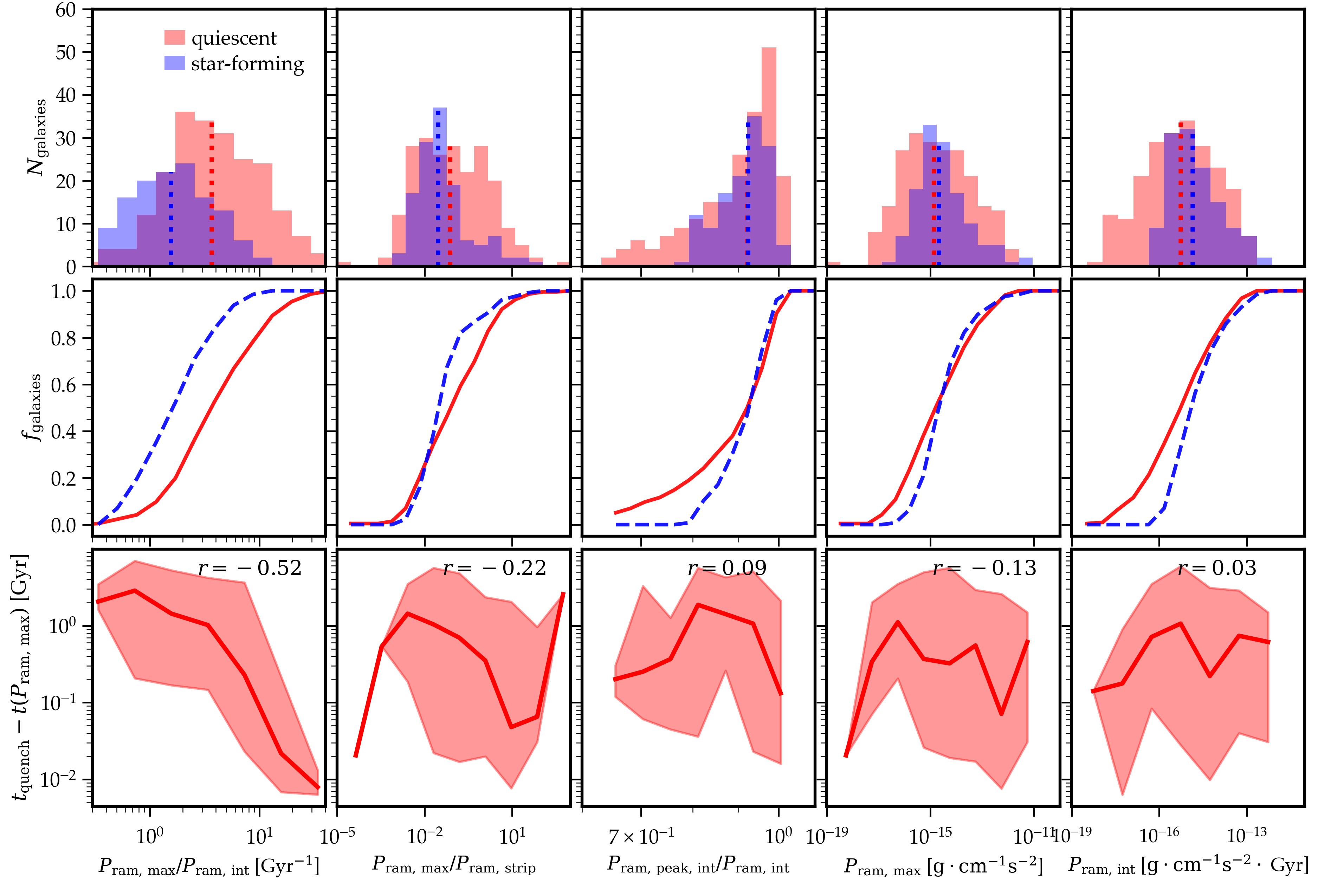}\par
\vspace{-4 mm}
    \caption{
    Samples (top), cumulative distributions (middle), and quenching time-scale correlations (bottom) for different ram pressure metrics.
    Dotted vertical lines in the top panels are the sample medians.
    The lines and shaded regions in the bottom panel are binned medians and 68 per cent scatter, respectively, for quiescent galaxies.
    The numbers printed in the bottom panels are the Pearson correlation coefficients for the full sample of quiescent galaxies.
    In the left column, we illustrate that our metric for ram pressure impulsiveness, $\impulsePram=\maxPram/\intPram$, separates the quiescent and star-forming galaxy distributions the most and correlates the strongest with quenching time-scale amongst the five ram pressure metrics.
    }
    \label{fig:metric_compare}
\end{figure*}

In Figure~\ref{fig:metric_compare}, we characterize our sample of quiescent and star-forming galaxies using five different ram pressure metrics; our preferred metric for ram pressure impulsiveness, $\impulsePram=\maxPram/\intPram$, is shown in the left column.
Impulsiveness is the only metric for which the sample medians for quiescent and star-forming galaxies are clearly offset; all of the other metrics have sample medians that are similar for the two sets of galaxies.
This fact is even clearer in the cumulative distribution functions (CDF) for each metric, where there is a clear separation of the quiescent and star-forming CDFs over the full range of $\impulsePram\sim10^{-1}-10^{2}\,\gyr^{-1}$.
Most importantly, impulsiveness is the only metric that has a significant correlation (Pearson $r=-0.52$) with quenching time-scale, the time between maximum ram pressure and last star formation.

The second from left column shows the same quantities for our simple quenching estimate metric, $\maxPram/\quenchPram$.
There are only marginal differences between the quiescent and star-forming distributions of this metric, and it has a weaker correlation with quenching time-scale.
The center column shows a new metric, 
\begin{equation}
P_\mathrm{ram,\,peak,\,int}=\sum_{i}{P_\mathrm{ram,\,peak}(t)\cdot\Delta t}/\intPram, 
\end{equation}
where the sum is done over snapshots where the ram pressure rises above the median ram pressure, which we denote as $P_\mathrm{ram,\,peak}$.
This metric can be thought of as an integrated, unitless analog of $\impulsePram=\maxPram/\intPram$.
Though the quiescent galaxy sample distribution of $P_\mathrm{ram,\,peak,\,integrated}$ has an extended tail towards lower values compared to the star-forming distribution, their medians are essentially the same.
This metric also does not show a correlation with quenching time-scale ($r=0.09$).

The two columns on the right show the distributions for the numerator and denominator of $\impulsePram$ separately.
Star-forming galaxies tend to have higher $\intPram$ because they have been integrated for longer and experience smaller ram pressure variations on average.
Neither $\maxPram$ nor $\intPram$ correlates strongly with quenching time-scale.

\bsp	
\label{lastpage}
\end{document}